\documentclass[10pt,a4paper]{article}
\usepackage{graphicx}
\usepackage{tabularx}
\usepackage{multicol}
\usepackage{tabto}
\usepackage{url}
\usepackage{enumitem}
\usepackage{textcomp}
\usepackage{eurosym}
\usepackage{wasysym}
\usepackage{xcolor}
\usepackage[labelfont=bf]{caption}
\usepackage[yyyymmdd]{datetime}

\usepackage{flowchart}
\usetikzlibrary{shapes,arrows.meta,chains,external}

\def\ccol{0.7em}

\newcolumntype{L}[1]{>{\raggedright\arraybackslash}p{#1}}
\newcolumntype{C}[1]{>{\centering\arraybackslash}p{#1}}
\newcolumntype{R}[1]{>{\raggedleft\arraybackslash}p{#1}}

\newcommand{\eat}[1]{}
\newcommand{\cz}[1]{\textit{\textbf{#1}}}

\newcommand{\ts}{
    \tikzset{>={Latex[width=3mm,length=3mm]}}
    \tikzstyle{line} = [draw, ->, >=latex, ultra thick]
    \tikzstyle{circ} = [
        circle,
        align=center,
        text width=2em,
        text centered,
        inner sep=1mm,
        outer sep=1mm,
        minimum width=0cm,
        minimum height=0cm
    ]
    \tikzstyle{block} = [
        rectangle,
        draw,
        text width=7em,
        text centered,
        rounded corners,
        align=center,
        minimum width=\smbwd,
        minimum height=1cm
    ]
    \tikzstyle{fdecision} = [
        diamond,
        draw,
        text width=6.5em,
        text centered,
        align=center,
        aspect=1.5,
        minimum height=0.5cm,
        minimum width=\smbwd,
    ]
    \tikzstyle{parallelogram} = [
        trapezium,
        draw,
        trapezium left angle=70,
        trapezium right angle=-70,
        text width=0.5em,
        text centered,
        align=center,
        minimum width=\smbwd,
        minimum height=1cm
    ]
    \tikzstyle{box} = [draw, rectangle, thick, fill=gray!30]
    \tikzstyle{boxy} = [
        draw,
        rectangle,
        thick,
        text centered,
        align=center,
        minimum width=4cm,
        minimum height=4cm
    ]
    \tikzstyle{noshape} = [text width=5em, text centered, minimum height=5em]
    \tikzstyle{nos} = [
        rectangle,
        text width=6em,
        text centered,
        align=center,
        minimum width=4cm,
        minimum height=1cm
    ]
}

\begin{document}

\bibliographystyle{plain}
\thispagestyle{empty}

\begin{center}
\large{\bf Digital Currency and Economic Crises: Helping States Respond}\\
\end{center}
\vspace{0.4em}
\begin{center}
\begin{minipage}[t][][t]{0.32\linewidth}
\begin{center}
\large{\bf Geoffrey Goodell}\\
\vspace{0.2em}
{\texttt{g.goodell@ucl.ac.uk}}
\end{center}
\end{minipage}
\begin{minipage}[t][][t]{0.32\linewidth}
\begin{center}
\large{\bf Hazem Danny Al-Nakib}\\
\vspace{0.2em}
{\texttt{h.nakib@cs.ucl.ac.uk}}
\end{center}
\end{minipage}
\begin{minipage}[t][][t]{0.32\linewidth}
\begin{center}
\large{\bf Paolo Tasca}\\
\vspace{0.2em}
{\texttt{p.tasca@ucl.ac.uk}}
\end{center}
\end{minipage}
\end{center}
\begin{center}
{\large University College London}\\
\vspace{0.4em}
{\large Centre for Blockchain Technologies}
\end{center}
\begin{center}
{\textit{This Version: \today}}\\
\end{center}

\begin{abstract}

The current crisis, at the time of writing, has had a profound impact on the
financial world, introducing the need for creative approaches to revitalising
the economy at the micro level as well as the macro level.  In this informal
analysis and design proposal, we describe how infrastructure for digital assets
can serve as a useful monetary and fiscal policy tool and an enabler of
existing tools in the future, particularly during crises, while aligning the
trajectory of financial technology innovation toward a brighter future.  We
propose an approach to digital currency that would allow people without banking
relationships to transact electronically and privately, including both internet
purchases and point-of-sale purchases that are required to be cashless.  We
also propose an approach to digital currency that would allow for more
efficient and transparent clearing and settlement, implementation of monetary
and fiscal policy, and management of systemic risk.  The digital currency could
be implemented as central bank digital currency (CBDC), or it could be issued
by the government and collateralised by public funds or Treasury assets.  Our
proposed architecture allows both manifestations and would be operated by banks
and other money services businesses, operating within a framework overseen by
government regulators.  We argue that now is the time for action to undertake
development of such a system, not only because of the current crisis but also
in anticipation of future crises resulting from geopolitical risks, the
continued globalisation of the digital economy, and the changing value and
risks that technology brings.

\end{abstract}

\section{Introduction}

The forced SARS-CoV-2 lockdown stopped a significant share of the real economy
activity across the entire world.  In retrospect, it might seem that the
decision to impose lockdown restrictions was straightforward.  However, it is
increasingly apparent that restoring the previous level of economic activity
will require some sort of stimulus, introducing the question of whether the
orthodox instruments of liquidity injection into the financial markets are the
right tools.  In this article, we consider whether an appropriately designed
digital currency would provide a useful alternative during financial and
economic downturns.  Specifically, we show how digital currency based on
distributed ledger technology (DLT) can:

\begin{itemize}

\item address the problem of resource distribution and allocation, and

\item reach the real economy directly, to support individuals and businesses
most in need of support.

\end{itemize}

We begin by introducing the following taxonomy of public digital currency.  On
the top level we define the Digital Value Container (DVC), a form of digital
asset infrastructure that includes all the forms of public digital currencies
issued by the State (either a central bank or the executive and legislative
branches of the government) that are part of the macroeconomic tools used to
manage or stimulate the economy.  DVC can further be segregated into:

\begin{itemize}

\item \textit{Central Bank Digital Currency} (CBDC), a fiat-based digital
currency issued by the central bank as part of the monetary policy toolbox; and

\item \textit{Fiscal Digital Currency} (FDC), an asset-based or fiscal-based
digital currency issued by the Treasury as part of the fiscal policy toolbox.

\end{itemize}

\subsection{The Challenge}
\label{ss:challenge}


We consider the varied and inconsistent nature of economic crises broadly,
using the current SARS-CoV-2 crisis as a touchstone.  We begin with the
consensus that SARS-CoV-2 crisis is not a financial crisis but a healthcare
system crisis driven by an exogenous shock and exacerbated by heightened levels
of public and private debt, as well as a labour market crisis in the real
economy resulting from the response to SARS-CoV-2.  We argue that any solution
that seeks to address the current situation by resorting \textit{only} to
injection of liquidity into the markets, as if the nature of the crisis were
essentially financial, will inevitably be weak and temporary.  Banks are
stronger now than 10 years ago.  Capital requirements are higher, regulation is
tighter, and the financial services industry continued to run without
discontinuity even during the lockdown.  We are not proposing a solution to a
pandemic, nor are we proposing that a digital currency is a proportionate
response to a pandemic.  Rather, we are suggesting that the pandemic
exacerbated certain underlying risks and ultimately led to policies that had a
particular effect on national and global economies. In light of this, we
propose a new tool for fiscal and monetary policymakers thinking beyond the
current crisis to future crises, which may be even more difficult to address.

In response to the 2008 US subprime credit crisis, the 2008-2009 Global
Financial Crisis, and the 2011 sovereign debt crisis, central banks injected a
great quantity of liquidity into the markets via unconventional monetary
policies which led to near-zero interest rates for commercial and household
loans. Indeed, OECD statistics show that aggregate leverage, as measured by the
debt-to-equity ratio of financial corporations, increased in 22 of 30 OECD
countries between 2006 and 2011~\cite{oecd2014}.  Paradoxically, the programs
used to solve the past financial crisis could turn out to be detrimental for
the current crisis.

Today's vulnerability does not lie mainly in the financial sector, but in the
wide range of indebted companies that will see revenues fall below critical
levels.  In this paper, we argue that DVC as CBDC or FDC can represent an
alternative, measured, and appropriate policy tool that also enables existing
policy mechanisms to target companies and people in need, to support long-term
continuity and stability.  Our proposal is set around creating the
infrastructure for cultivating economic activity and exchange in all
circumstances while maintaining the privacy of users.


Some risks are impossible to detect or even imagine in advance. Disasters such
as Hurricane Katrina in 2005, the Deepwater Horizon oil spill in 2010 and the
Fukushima nuclear accident in 2011 were unexpected and extremely damaging.
Likewise, SARS-CoV-2 belongs to this category of risks that are
unknown-unknown: shocks that are unidentified and uncertain, and which might be
a consequence or exacerbator of recent policy or existing risks.  Who would
ever imagine that a virus possibly triggered in a live-animal market of Wuhan
would have locked-down entire continents for months?  The challenge we face is
not only the shock itself but the reaction to the shock.  The major obstacle to
addressing SARS-CoV-2 is the endogenous risk amplification mechanism that will
derive from actions taken as the shock gathered momentum from the endogenous
response of people, firms, and governments.  The lockdown is damaging the
overall economy, and projections for 2020 by the IMF are that both advanced and
emerging economies will be in recession with a growth level of -6.1\% for the
advanced economies and -1.0\% for emerging and developing
economies.\footnote{The Federal Reserve Bank of New York Nowcast Model expects
US GDP to decrease by over 30\% in the second quarter of
2020~\cite{b1}.}~\cite{gopinath2020}

The current shock to the global economy, exacerbated by the SARS-CoV-2 pandemic
is more severe, faster, and more multi-faceted than both the Global Financial
Crisis and the Great Depression. In both prior periods, unemployment rates
soared above 10\%, annualized GDP contracted significantly due to deflation,
insolvencies skyrocketed, and stock markets collapsed. While those scenarios
played out over three years in the previous crises, the current crisis
demonstrated similar attributes in a few short weeks~\cite{rushe2020}.

Central banks of several countries have maintained a zero or near-zero policy
rate to date, and governments around the world have announced (and as of today,
partially implemented) massive fiscal stimulus measures to stimulate the
economy.  Several central banks and governments have recently announced new
measures of monetary and fiscal policy, ranging from quantitative easing and
various types of asset purchase programmes, such as £200 billion by the Bank of
England~\cite{b10} and over \euro750 billion by the EU~\cite{b11}.  By
\textit{quantitative easing} (QE), we refer to the central bank conducting
asset purchases by printing new money from the private market that is mediated
by banks, who then interact with non-banks, and we mean to implicitly include
all of its various forms.\footnote{For example, selling short-term securities
and using the proceeds to purchase long-term securities.}  Central banks have
also cut interest rates that were already low, such as the Bank of England
reducing its policy rate by 0.65\% to 0.1\%~\cite{b12} and the US Federal
Reserve reducing its policy rate to 0-0.25\% from 1-1.25\%~\cite{b13}, and
governments have implemented direct fiscal stimulus programmes, including over
\$2 trillion in the US and over \$15tn across the G10 countries~\cite{b14},
each with the aim of stimulating the economy~\cite{b15}.

The parts of the economy that will suffer the most are characterised by
millions of self-employed persons as well as small and medium-sized companies,
particularly in the tourism and service industries, that will have difficulty
remaining in the market, especially in countries with large informalities.
Those areas of the economy are more disadvantaged and unprotected as they will
not have access to the funds channelled through open market operations led by
central banks, and also because they will face gatekeepers, bureaucracy, and
other obstacles to accessing funds and financial support imposed by the banking
sector.  The result is a set of gaps in the global economy that are not covered
by existing response mechanisms.

To preserve those areas of the real economy, such market participants will need
a fast and efficient transfer of in-kind and cash contributions.  For this
reason, we acknowledge the growing literature on CBDC and extend it by
emphasising the potential role of FDC as new policy tool that could be
introduced into the toolkit of treasury departments to better support economic,
social, and environmental objectives, especially in times of crisis.  In
contrast to traditional fiscal policy and monetary policy transmission
mechanisms, DVC based on distributed ledger technology can offer almost
cost-free instant cash transfers where it is needed.  A particularly
interesting innovation is that DVC can enable programmable money, which
introduces a series of features that make DVC a valuable alternative policy
tool and payment infrastructure. In line with the above observations, we
propose a DLT-enabled digital currency system which has been argued to deliver
a variety of economic and operational benefits~\cite{bis2018}.  In this
article, we mainly focus on CBDC rather than FDC,\footnote{Or other possible
assets, although many of our points are transferrable.} highlighting the most
important characteristics of our approach, with the rest being left to
policymakers whose objectives are specific to context, timing, and location.

\subsection{The Solution Landscape}
\label{ss:landscape}

IMF research by Tommaso Mancini-Griffoli and others identified a tension in the
potential design features of a CBDC~\cite{mancini2018}, which we recast and
sharpen here as a trilemma involving scalability, control, and privacy, of
which not all three can be fully achieved at the same time in the context of
private ownership and use of money.  Bank accounts have near-perfect
scalability and control at the expense of privacy.  Cash has privacy and a
measure of control that limits its scalability.  It is difficult to imagine a
system with perfect control because it would result in real ownership being
meaningless and because there will always be some malfeasance in use. The same
is true with perfect privacy because there will always be software bugs and
timing attacks as well as limited practical uses, whereas perfect scalability
would mean that sufficiently large transactions are not endogenously meaningful
or that the system does not truly serve the public interest.

Mancini-Griffoli and his co-authors are right in their assessment that
anonymity is an important feature of cash, that privacy of transactions is
essential, and that the specific design features of CBDC could have a
significant impact on financial integrity~\cite{mancini2018}.  Our proposal
provides a solution with the flexibility to accommodate the widely-acknowledged
requirements and goals of CBDC and which is more akin to cash.  Specifically,
it delivers a measure of control by restricting peer-to-peer transactions.
However, it does not offer the near-total degree of control that seems to be
taken as a requirement in some designs~\cite{auer2020}, and instead its retail
applications are exposed to a corresponding limitation to their scalability,
but not one that cannot be overcome by introducing additional control, in
limited contexts, outside the operating plane of the ledger.  Our system
provides a model for modulating the degree of control, allowing government
actors to finely tune their choice of trade-offs in the trilemma.  For example,
it might require that certain (or all) businesses cannot accept payments larger
than a certain size without collecting or reporting additional information that
limits privacy, or it might require that some individuals or non-financial
businesses have a larger or smaller cap on the volume of their withdrawals into
private wallets.  To draw an analogy, it operates like an automated conveyor
belt holding keys that are trying to meet a lock, and if they are the right
fit, as determined either at large or on a case-by-case basis, then the
transactions take place in an automated way.

It is the intrinsic design of our proposal that ensures privacy for its
transactions; our design seeks to be private \textit{by default}.  We do not
envision privacy as something that can be bolted on to a fully-traceable system
(for example, with ``anonymity vouchers''~\cite{r3-cbdc,dgen}) or that can
depend upon the security or protection offered by some third party.
Conversely, the features that apply on a case-by-case basis, such as limits to
the size of withdrawals to anonymous destinations or limits to the size of
remittances into accounts from private sources, that are external to the core
architecture and can be managed by policy.

In May 2020, Yves Mersch, Vice-Chair of the Supervisory Board and Member of the
Executive Board of the European Central Bank, acknowledged the importance and
significance of preserving privacy, suggesting that an attempt to reduce the
privacy of payments would ``inevitably raise social, political and legal
issues''~\cite{mersch2020}.

This is important for three reasons.  First, no digital currency, token-based
or otherwise, would guarantee complete anonymity: consider the potential for
timing attacks and software bugs.  Even bank notes do not achieve perfect
anonymity; consider their serial numbers, and the possibility wherein
individual notes can be marked.  Nevertheless, we must consider the
implications of systems that attempt to force users into payment systems with
different anonymity properties and trade-offs in general.  Second, we have an
opportunity to demonstrate a system that can achieve and deliver a measure of
true privacy, in contrast to an assumption that there must be exceptional
access, or that privacy is not the starting point but rather something that
should be protected by an authority~\cite{benaloh2018}.  Such a system, which
we describe in Section~\ref{s:design}, would constitute an improvement over
both the various institutionally supportable digital currency systems that have
been proposed to date as well as the various ``outside solutions'' involving
\textit{permissionless} ledgers that are used in cryptocurrencies such as Zcash
and Monero.  Third, it demonstrates that privacy is sufficiently important that
we should not rush headlong into creating infrastructure, or allowing
infrastructure to be created, that might forcibly undermine it.  In contrast to
\textit{data protection}, which is about preventing unauthorised use of data
following collection, \textit{privacy} is about preventing individuals (and in
some cases businesses) from revealing information about their (legitimate)
habits and behaviours in the first instance.  As an architectural property,
therefore, privacy is a fundamental design feature that cannot be ``granted''
or ``guaranteed'' by some authority.

In the same statement, Mersch also stressed the importance of the role of the
private sector in operating a network for payments:

\begin{quote}
``[D]isintermediation would be economically inefficient and legally untenable.
The EU Treaty provides for the ECB to operate in an open market economy,
essentially reflecting a policy choice in favour of decentralised market
decisions on the optimal allocation of resources. Historical cases of
economy-wide resource allocation by central banks are hardly models of
efficiency or good service. Furthermore, a retail CBDC would create a
disproportionate concentration of power in the central
bank.''~\cite{mersch2020}
\end{quote}

A few months before Mersch's speech, Tao Zhang, Deputy Managing Director of the
International Monetary Fund, also offered his opinion on the current set of
proposals for CBDC, which he said ``imply costs and risks to the central
bank''~\cite{zhang2020}.  We argue that his conclusions follow from the
proposals that have been elaborated so far by central banks, which have
generally involved a central ledger operated by the central bank
itself~\cite{mas2018,boe2020}.  We suggest that such proposals have been
designed neither to be holistic nor to complement the current model of
payments, settlement, and clearing that exists today.  In contrast, our
approach specifically avoids the costs and risks identified by Mersch and
Zhang, which we characterise more specifically in Section~\ref{ss:dlt}, and is
broadly complementary to the current system.

Zhang also introduced the idea of a ``synthetic CBDC'' consisting of tokens
issued by private-sector banks~\cite{zhang2020}.  We argue that the desirable
qualities that Zhang ascribes to synthetic CBDC apply to our proposed solution
as well, except that our proposed solution still allows for ``real'' CBDC and
other assets issued directly by the government (such as FDC), although the
infrastructure would be operated by private-sector \textit{money services
businesses} (MSBs), which we shall describe in Section~\ref{s:policy} and for
our purposes comprise both traditional commercial banks and financial
institutions as well as new entities that would only have central bank reserves
as their assets and whose liabilities would in turn only be deposits.  This is
an important distinction, and although Zhang provides no specific description
of the technical features of synthetic CBDC, we assume that it would not
involve a distributed ledger and that it would not be possible to have private
transactions, since the private-sector banks would have visibility into the
operation and ownership of their own tokens.

Finally, we argue that our approach must be token-based, with accounts used
peripherally to the token infrastructure and with key design features
integrated into the system itself.  We specifically disagree with the argument
of Bordo and Levin favouring the use of accounts~\cite{16}.  Specifically, the
trust property we seek is intrinsic to the token itself, not to any specific
account-granting institution or system operator.  We also explicitly state:
\textit{Trust cannot be manufactured and must be earned}.  More importantly, we
do not create trust by asking for it; we create trust by showing that it is not
needed.  The approach that we describe in Section~\ref{s:design} addresses this
requirement directly.

\section{DVC as a Policy Tool}
\label{s:policy}

In 2020, local and global, non-digital and digital economies face a milieu of
risks preconditioned by the growing levels in sovereign public and private
accumulation of debt including liquidity traps within the economy and safety
traps related to assets.  It is further accentuated by the growing gap between
the financial economy and the real economy, the reasons for which we shall not
explore here, thus indicating a lacuna in the toolkit of monetary and fiscal
policymakers in being able to bridge that gap, particularly where, for example,
deposit rates (and rates in general) do not frequently and speedily respond to
monetary policy, nor do the behaviour of households and firms~\cite{3}. Private
household debt in the US is higher than in 2008~\cite{4}, and in the UK it is
15\% lower whereas household unsecured debt is at an all-time high~\cite{5}. We
need to \textit{strengthen the monetary and fiscal policy transmission
mechanisms} with economic stimulus and financial stability mechanisms that are
able to reach the real economy.

In Section \ref{s:FiscMon} we shall argue why and how DVC, in all its
derivations as CBDC and FDC, represents an innovative underlying infrastructure
that can further enrich the economic stimulus of both monetary and fiscal
policies. In the second part \ref{s:demandCBDC} will continue by stressing on
the role and demand for CBDC and Section \ref{s:design} will propose a new
model design for CBDC and FDC.

\subsection{Fiscal and Monetary Policy Transmission Mechanisms}
\label{s:FiscMon}

DVC, as CBDC or FDC, represents a genuine risk-free asset that is capable of
absorbing macroeconomic shocks and preserve its economic value that could be at
the disposal of central banks and governments~\cite{6}. Governments and central
banks would have a variety of tools available to them to manage their DVC and
its impact and be capable of effecting quantity limits, implementing active
pricing and collateralising them with sovereign bonds and other
assets~\cite{7}. This is in addition to creating truly efficient payment
infrastructure, which, in combination, can enhance the resilience of the
financial system, increase consumer confidence, stimulate the overall
economy~\cite{8}, and facilitate greater economic interactions with less
friction, particularly during times of economic difficulty~\cite{9}.

\paragraph{Fiscal Policy Transmission Mechanisms.}
As highlighted in Section~\ref{ss:challenge}, the post-SARS-CoV-2 world economy
will inevitably fall into recession. During recessions, post-Keynesian
economists promote the direct intervention of the State in the economy to
encourage and sustain the private sector and internal consumption
\cite{davidson2011post}. Historically, the mechanism of economic stimulus split
into monetary and fiscal stimulus.

Monetary policy is controlled by the central bank which primarily uses tools
such as modulating the discount rate, performing open market operations and QE,
and adjusting the reserve ratio.  In the case of a real-economy shock, monetary
stimulus attempts to increase the amount of money and credit in the economy to
boost consumption. These mechanisms generally rely upon the traditional banking
and payment sectors as channels of transmission of credit or liquidity
stimulus.  However, in times of shocks to the real economy, the overall
expansion of the money supply to influence inflation and growth may have less
impact on the real economy.  Since the monetary transmission mechanisms mostly
rely upon the channel of bank credit to firms, tighter credit supply impairs
the transmission of monetary policy to the real economy~\cite{pt1}.  No one can
prevent a solvent bank from refusing to lend to firms.  Indeed, a weakly
capitalised bank may improve its solvency metric by cutting credit exposures.
Concerning this, the asset purchase programs by central banks of private
non-financial securities has been designed to unblock the transmission of
accommodating financial conditions through banks to ultimate
borrowers~\cite{pt2,pt3}.  However, those asset purchase programs are generally
designed for listed securities which respect the designated selection criteria.
For example, the Pandemic Emergency Purchase Program (PEPP) recently put in
place by the ECB as a further response to the SARS-CoV-2 crisis sets stringent
minimum rating requirements for corporate bonds purchase~\cite{pt4}.

As a matter of fact, small and medium-sized enterprises (SMEs, i.e. firms with
fewer than 250 employees), self-employed individuals, and households are cut
out from these programmes.  This is an issue for countries populated by SMEs
which, due to their narrow equity base and limited access to credit markets,
meet difficulties in finding appropriate financing~\cite{pt5}.\footnote{In
Europe there are 25m SMEs which employ over 90m people. According to smeunited,
a European lobby group, 90\% of Europe's SMEs are affected by the recent shock
and 30\% of them say they are losing 80\% of sales or more~\cite{pt6}.}

This impairment combined with the fact that income inequality may increase
during economic crises \cite{de2012earnings}, alerts policy makers who
therefore have designed alternative channels of liquidity transmission to
support small firms and people living under income stress and below the limits
of poverty.  One possible solution to this financing gap problem has been the
introduction of Credit Guarantee Schemes (CGSes) which are designed to help
banks absorbing more risk from the real economy by supporting lending to
low-quality borrowers~\cite{pt7,pt8}.
CGSes allow the partial transfer of credit risk stemming from a loan or a
portfolio of loans. The goal is to close the financing gap by substituting
collateral provided by a borrower with credit protection provided by an
external guarantor, the State.  In this respect, they show similarity to credit
insurance products and credit default swaps. Still, the final lending decision
stays with the bank, a market-based, private-sector entity that is assumed to
have the expertise and technology necessary to evaluate credit applications and
projects~\cite{vi2014}.

Additional schemes instituted by governments to sustain  self-employed
individuals  and households are  Guaranteed National Income schemes, Minimum
Income schemes or the Universal Income schemes. One such instrument is the
``social card'': a prepaid electronic payment card, generally issued by the
ministry of economic development or work and pension office and their
affiliates~\cite{pt9}.  These social cards are generally deployed on top of
payment networks such as Mastercard or VISA, and their use is limited only to
certain authorised POS shops like groceries and food stores.  Therefore, users
are required to undergo the same pre-authentication and authorization
procedures that apply to any other debit or credit card.

However, evidence from the market suggests that these mechanisms do not always
yield the expected results~\cite{pt10}. Only carefully designed and
continuously evaluated products have a chance to deliver the associated public
policy objectives.  Without this type of care, such mechanisms can do more harm
than good by misallocating resources to companies and citizens, by providing
credit/liquidity where is not needed, or by rationing credit or liquidity where
it is most needed, thus crowding out private collateral and unnecessarily
increasing public debt.

Additionally, theoretical and empirical studies raise concerns about the
ability for CGSes to sustain the economy~\cite{gfdr2013}.  In particular, CGSes
are considered expensive tools that pose problems of financial sustainability.
At the same time, the benefits have yet to be proven, as there is no conclusive
evidence that they allow additional loans to SMEs with financial
restrictions~\cite{tanner2015,tanner2015a}.  Indeed, economic crises are often
accompanied by a flight-to-quality by lenders and investors, as they seek less
risky investments, frequently at the expense of SMEs~\cite{ferri1998},
primarily because the credit rationing hypothesis is exacerbated
\cite{jaffee1976imperfect}, and \cite{stiglitz1981credit}.

We conclude arguing that there are important differences between the current
crisis and the 2008 financial crisis~\cite{pt11}, and when the target is not
the financial sector but the real economy, the banking channel can turn out to
be a second-best choice and ultimately a bottleneck.

Thus, without entering into the debate on the origins and the basis of the
``State entrepreneur'' who directly intervenes to support the economy, we
should ask whether new channels of financial stimulus transmissions can be
designed as alternatives to traditional banking and payment transmission
channels.  Concerning this, we believe that it is possible to implement new
types of fiscal policy transmission mechanisms in the form of FDC to better
fulfil the goals of the executive and legislative branches of government.

We argue that an FDC-based payment infrastructure would give the State a more
direct control and understanding of the economic system under distress. Such
control would permit better intervention in response to any exogenous shock and
business cycle while also ensuring better individual compliance with tax
collection and anti-money laundering statutes~\cite{raskin2016}.  Our proposed
design in Section~\ref{s:design} introduces a general DVC design concept that
can be adjusted and turned into a government-issued electronic token that can
be used to inject non-fiat based liquidity into the real economy. One possible
form of liquidity could be for example represented by a tax discount (similar
to the tax concessions that are generally given for example to the construction
industry) to all citizens and all SMEs. In this particular case, the FDC could
become a kind of electronic multi-year certificate whose coupons at maturity
can be used to pay the national taxes over time.  This is only an example; the
FDC is a neutral value container that, based on the value or claims that it
represents and its technical implementation features, it can fulfill, without
using the banking transmission mechanism, one of the following four budgetary
models: (1) remunerated and repayable loan; (2) remunerated and non-repayable
loan; (3) non-remunerated and repayable loan; (4) outright grant.

Because our solution is based on a DLT system, it can help to close the
financing gap by substituting State-sponsored credit guarantees with public
co-funding schemes directly targeting companies, SMEs and also people in need.
By doing so we expect to achieve:

\begin{enumerate}

\item Elimination of the double asymmetry of information between State and
bank, and between bank and borrower;

\item Elimination of credit rationing;

\item Lower funding costs;

\item Shorter liquidity-injection time.

\end{enumerate}

The FDC could become an instrument to revive the strength of the traditional
fiscal policy and overcome some of its traditional drawbacks~\cite{pt12}.  More
generally, the FDC can be seen as a direct channel for the transmission of
fiscal stimulus shocks into the economy, particularly helpful in cases wherein
the efficacy and feasibility of other tools runs thin. This could be balanced
by the use of CBDC as a new monetary policy instrument, as we shall discuss
later in this section.  We therefore think that especially in periods of crisis
and downturn, State-controlled digital currencies can become an important part
of the counter-cyclical public policy toolkit~\cite{15}.

\paragraph{Monetary Policy Transmission Mechanisms.}

The monetary policy transmission mechanism refers to the process and efficacy
by which the policy rate influences inflation and the real economy and
comprises two main parts: (i) the pass through of the policy rate to other
rates in the economy including the money market rate, lending and deposit
rates, and (ii) how and to what extent these rates influence inflation and
reach the real economy.

There are presently two main forms of monetary policy; (i) traditional monetary
policy; and (ii) unconventional monetary policy. Traditional monetary policy
focuses on changing the short-term interest rate on reserves in order to affect
changes to the real interest rate in the economy and meet the inflationary
target. Unconventional monetary policy, which became prevalent during and in
the aftermath of the Global Financial Crisis includes several variations, most
notably QE.  Other types include setting negative interest rates on reserves in
order to affect long-term interest rates within the economy, forward guidance
to manage expectations within the economy such that rates are viewed as being
consistent in the long-run and affecting behaviour accordingly, as well as
other forms of QE, such as yield curve control (which has not been effectively
implemented outside of Japan since the 1940’s, yet there is growing and renewed
interest~\cite{a4}).  It is safe to say, however, that the overnight interest
rate, QE, and forward guidance and various forms of asset holdings are
permanent forms of monetary policy~\cite{a5}.

In particular, concerning the reserve or overnight interest rates in several
major economies, two major problems have been highlighted: (1) the zero-lower
bound~\cite{defiore2018} and (2) the fact that rates are not always passed to
market rates effectively, such as those of private borrowers~\cite{a6}.  With
respect to the extension of QE measures, the risks that have been highlighted pertain
to the excessive expansion of the central bank's balance sheet that may lead to
deep financial recessions in the future and render other unconventional
monetary policy tools ineffective~\cite{a7}.  Furthermore, the direct impact of
QE is relatively small. For example, the median estimate of a bond purchase of
approximately 10 percent of GDP results in eight tenths of a point in reduction
of long-term interest rates~\cite{a8}.  Forward Guidance on the other hand,
rests on the assumption that effectively communicating policy changes, or the
lack thereof in the future, will affect economic decisions is
somewhat at odds with the reality of today's world~\cite{a5}.  Neither QE nor
forward guidance, broadly speaking, although initially construed in their
various forms to generate private sector spending, were able to stimulate
significant and long term private sector borrowing~\cite{a5}.

Although we do not question that all of these tools can be useful in addressing
the need for stimulus, we recognise that the current environment, characterised
by low interest rates and growing debt (both public and private), as well as
the current crisis of the real economy resulting from an exogenous shock from
outside the financial industry, together present a singular context both unlike
those in which such tools have been previously applied and characteristic of
potential future environments.  For this reason, rather than assuming that the
universe of potential methods for effective stimulus is limited by innovations
of the past, we consider a broader range of approaches, starting with the
question of how we can most effectively reach the real economy, particularly
the ascendant digital economy, through CBDC using our DVC approach.

In this context, we see the potential of CBDC as an additional monetary policy
transmission tool.  In particular, the ongoing discussion in the community
revolves around the potential of CBDC as: (1) a countercyclical tool; (2)
``helicopter money''; (3) an interest-bearing instrument; (4) a floor-system
instrument; and (5) a QE instrument.  We consider each in turn:

\begin{enumerate}

\item Several economists are now convinced that CBDC would simplify the options
available~\cite{13} as well as broaden the availability of options provided to
central banks in their monetary policy toolkit. It can also be seen to increase
welfare if it were to reduce banks' deposit market power~\cite{14}.  It has
also been shown that were the issuance of CBDC to be implemented by means of
purchasing government bonds equivalent to 30\% of GDP~\cite{15}, it could lead
to an increase in GDP of 3\% over two decades, while maintaining price
stability given that the real value of the CBDC could be held over
time~\cite{16}. Furthermore, it is possible for the issuance and control over
the supply of a CBDC to have a stabilising effect on the business cycle if
adopted countercyclically~\cite{15}.  Namely, this is the case if CBDC would be
interest bearing, or enables functionality, such as through taxes and subsidies
that allow it to mimic the effects of remuneration.  Thus, CBDC would become a
policy instrument that would allow for new tools to improve the effectiveness
and implementation of monetary policy, such as through the stabilisation of the
business cycle by being able to control the issuance of the CBDC and the
countercyclicality of its price in response to economic shocks to private money
creation or demand~\cite{15}, as well as equip central banks with the ability
to establish price stability targets\footnote{We imagine this would be done by
facilitating the collection of data at a technical level.} as opposed to
inflation targets, if they so choose~\cite{16}.

\item As we discussed previously, DVC could provide for new forms of
money-financed fiscal programs through targeted ``helicopter money''~\cite{22}
in the sense that FDC could increase cash-like holdings within an economy by
making direct lumpsum issuances to individuals, businesses and households,
particularly in instances of escaping a liquidity trap and lessening the
possibility of deflation, or in the event that monetary transmission mechanisms
see little result or efficacy, particularly when, at present, interest rates
have already been at longstanding lows~\cite{23}. This is particularly relevant
in periods of economic uncertainty or a looming economic crises in the ability
to capitalise the economy with outside money by means of new, novel and more
direct channels.  DVC would allow States including central banks to provide
liquidity much faster in the case wherein cash is increased\footnote{Since the
2008 global financial crisis, the cash to GDP ratio has increased worldwide.
This may be caused by the need for a store of value, or alternatively payment
needs or transaction demand~\cite{34}  In this context, interest rate policy
may be constrained by the availability of cash because as interest rates fall
below zero, the demand for cash increases and a shift to cash ensues, notably
with a cost associated with it~\cite{35}.} and deal with risks in a much more
direct, timely and targeted way, and increase reserves
electronically~\cite{36}.  In summary, the central bank would be capable of
theoretically increasing or decreasing the supply of outside money in the real
economy through direct digital channels~\cite{24}.

\item Another ongoing discussion in the community is whether CBDC should be an
interest-bearing instrument.  We think that an interest-bearing CBDC would make
pass-through of rates more direct. For example, if CBDC were to be convertible
to commercial bank deposits as we propose in our model (including for cash as a
matter of design), banks would have less control in setting the interest rate
on deposits.  Therefore, a change in the policy rate could be transmitted much
more directly to bank depositors. This would be done instead of more
conventional alternatives such as bills issued by the central bank, reverse
repo facilities and time deposits as liquidity absorbing instruments. Relative
to today, the existence of the CBDC and its potential remuneration could
strengthen the pass-through to deposit rates because commercial banks would
want deposits to be at least as attractive as CBDC~\cite{armelius2018}.  We do
not, however, necessarily propose that interest rates would be paid to holders
of digital currency; instead, we propose features that replicate it using
(dis)incentives, vintages, taxes, and subsidies directly on the asset itself,
as we shall discuss in Section~\ref{s:analysis}.

\item Let us also consider the potential role of CBDC in the case that monetary
policies are set under a floor system. Since the Global Financial Crisis,
central banks have increased their reserves, primarily through a floor system
whereby reserves are expanded by asset purchases to clear the market at the
rate paid on those reserve balances, and whereby the policy rate is set on
reserves, and not on excess reserves, which are often different. Different
positions are taken as to whether the interest rate set on reserves is binding,
we shall assume that due to the fact that banking currently operates in a
tiered model, access to central bank money is limited due to the frictions in
place. With a CBDC and MSBs holding central bank money as reserves with the
central bank, this could be made more binding.

It would also be possible to expand the quantity of CBDC so that the market
clears at the floor. The rate could then be used on CBDC balances to guide
rates in the rest of the economy, \textit{pro tanto} both quantity and price of
CBDC could theoretically be varied to stimulate the economy through aggregate
demand. Aggregate quantity could also be expanded as a method to provide
economic stimulus to the economy in a more direct and targeted way because both
the price and quantity can be varied in a floor system~\cite{29}. This would
similarly apply to the operation of a corridor system where supply of CBDC is
reduced to create a secondary market for CBDC that clears above the rate of
reserves and which could operate as an intraday market and allow for more rapid
and direct monetary policy~\cite{30}.

The model outlined in Section \ref{s:design} addresses this discussion by
allowing also MSBs, including but not limited to banks and non-financial
institutions, to buy or lend within the CBDC market to trade central bank
money, thus increasing liquidity, although this would remain separate from
reserves~\cite{30}. Such a CBDC would become a liquid and creditworthy
financial instrument.  It would be similar to interest bearing central bank
reserves or reverse repo facilities, although it would be tradeable and
accessible by the general public (and potentially remunerated using specific
features).  We imagine that in practice, it could operate more similarly to
short-term maturity government bills that are programmable and which have
several variants available at any given time.

\item  The overall goal of CBDC as QE instrument would be to sustain or
increase the GDP \cite{15}. It is important to note that the issuance of CBDC
entails that the funds need to be invested by means of asset
purchases~\cite{26}. A CBDC could allow MSBs that hold a reserve account with
the central bank to sell assets directly to the central bank.  It can therefore
be much more efficient and more developed as it relates to the real economy.
For example, the central bank may be able to better identify the types of
business sectors for which the balance sheets of the MSBs that serve such
businesses may be altered, thus strengthening the impact of the transmission of
QE to the real economy.

\end{enumerate}

\subsection{The Demand for CBDC}
\label{s:demandCBDC}

Demand for CBDC will certainly be a function of various design features and
their policy objectives and ultimately determined by household preferences for
payment instruments. It will be determined by, on the one hand, the features
that are similar to either cash or bank deposits, and on the other novel design
features that might or might not replicate a particular effect of a traditional
money instrument.  For example, the closer a CBDC is to cash, the more likely
it will be seen as a substitute to cash, similarly with bank deposits.
Ultimately, the payment choice and demand for CBDC by households will be
influenced by how near they are to either or both, and how well it offsets
their inherent unchangeable design qualities.

Whether CBDC results in great demand and demonstrates itself to be an
attractive asset to hold as a liquid, credit risk-free, and price stable
central bank liability will ultimately be determined by the policy objectives
and the tools used as it relates to decisions on the following:

\begin{enumerate}

\item access,

\item availability,

\item remuneration and (dis)incentives, and

\item ancillary design features that may be inherent but used only
for specific objectives.

\end{enumerate}

In addition, were a CBDC designed not to provide certain qualities of privacy,
some users would remain avidly dedicated to the use of cash~\cite{agur2019}.
Our proposal, described
in Section~\ref{s:design}, disrupts this notion and shows how a measure of true
anonymity can be maintained.  A CBDC could support replacing private sector
assets into risk free assets to address the safe asset shortage, particularly
given that although bank deposits are broadly insured up to some amount, they
continue to exhibit credit and residual liquidity risks.  Moreover, there is
demand for semi-anonymous means of payment~\cite{20}, as well as for a variety
of instruments capable of being used for payment, and due to heterogeneity in
the preferences of households the use of a CBDC has immediate social
value~\cite{21}, both of which are direct consequences of our proposal.

Our proposal frames CBDC as a distinct financial instrument but one that
nonetheless shares many features with cash, including being fully
collateralised and not providing for the ability to lend or rehypothecate.
Moreover, we are not proposing the abolishment of bank notes, nor of bank
deposits. On the contrary, we understand all three instruments to have merit
and value to households and firms within an economy and can be used to
complement one another and increase the overall welfare of individuals and
firms through the adoption of CBDC~\cite{21}.  An example of the inherent
difficulties within proposals that argue for the abolition of cash is that the
increase in its use is predominantly situated within lower socioeconomic
segments of a community, and using CBDC to drive out cash would adversely
impact those households and firms.

Furthermore, the CBDC proposed in our design model relies upon the DLT
infrastructure for a variety of reasons outlined in Section \ref{s:design}.  In
our view, this is currently the most plausible method of implementation whereby
the central bank can collaborate with private sector firms, via either private
public-private partnerships or other collaborative and supervisory models, to
deliver a national payments infrastructure operated by the private sector.  The
use of DLT does not imply that households and retail members of the public must
have a direct account or relationship with the central bank, as wrongly assumed
by some.  On the contrary, our design recognises the important role of MSBs,
especially for identifying, onboarding, and registering new customers,
satisfying compliance requirements, and managing their accounts (if
applicable).

MSBs do not necessarily perform all of the functions of banks, such as lending
credit.  Moreover, in our design, we envisage full convertibility at par across
CBDC, bank deposits, bank notes, and (for authorised MSBs) reserves, both to
ease its introduction and to not interfere with the fungibility and general
composition of the monetary base.  To whatever extent this involves limitations
or the introduction of frictions will be a matter of policy.  Yet, in
principle, at-par convertibility for cash and bank deposits as the default is a
practical and design necessity.  Issuing and introducing CBDC enables a new
policy tool in adjusting the (dis)incentives to hold the CBDC through its
various features but also to balance the possible flight from bank
deposits~\cite{10}, for which we do not see CBDC as a general substitute.

\section{Our Proposal}
\label{s:design}

The core of our proposed design is based upon an article by Goodell and
Aste~\cite{goodell2019}, which describes two approaches to facilitate
institutional support for digital currency.  We build upon on the second
approach, \textit{institutionally-mediated private value exchange}, which is
designed to be operated wholly by regulated institutions and has the following
design features:

\begin{enumerate}

\item Provides a \textit{government-issued electronic token} that can be used
to exchange value without the need for pairwise account reconciliation.

\item Allows transaction infrastructure (payments, settlement, and clearing) to
be operated by \textit{independent, private actors}\footnote{Presumably, the
independent, private actors would participate in the activities of a
co-regulated authority, such as FINRA in the United States, or a quango, such
as FCA in the United Kingdom.} while allowing central banks to control monetary
policy and CBDC issuance, with control over the creation of CBDC but not its
distribution.  (The same applies to government in the case of FDC.)

\item Protects the \textit{transaction metadata} linking individual CBDC (or
FDC) users to their transaction history by design, without relying upon trusted
third parties.

\item Affords regulators \textit{visibility} (excluding counterparty
information) into every transaction, allowing for analysis of systemic risks.

\end{enumerate}

In this section we describe the central assumptions underlying our proposal,
and we identify the benefits of distributed ledger technology (DLT) and offer
support for our claim that a DLT-based architecture is necessary.  Then, we
describe how our proposed mechanism for digital currency works at a system
level, identifying essential interfaces between the institutional and technical
aspects of the architecture.  We conclude by explaining how we would leverage
our proposed architecture to achieve the economic stimulus objectives of State
actors and to facilitate payments by individuals and businesses.

\subsection{Key Assumptions}
\label{ss:assumptions}

We imagine that digital currency might be issued by a central bank as ``true''
\textit{central bank digital currency} (CBDC), or it might be issued by
government as FDC, representing an obligation on a collateralised collection of
State assets, such as sovereign wealth or Treasury assets.  In either case, we
note that in many countries (including the UK), no single party (including the
central bank) has been assigned the responsibility to design, maintain, and
update the rules of the process by which financial remittances are recorded and
to adjudicate disputes concerning the veracity of financial remittances.  We
also note that responsibility to operate transaction infrastructure and
supervise payment systems is different from the responsibility to create tokens
and safeguard the value of State currency.  In many countries, systems for
payments, clearing, and settlement are a collaborative
effort~\cite{bis2012,bis2012a}.  A design that externalises responsibility for
the operation of a transaction infrastructure supporting digital currency is
not incompatible with the operational role of a central bank in using digital
currency to create money and implement monetary policy.

Our approach to digital currency differs substantively from the vision proposed
by several central banks~\cite{mas2018,boe2020}.  We argue that the purpose of
digital currency is to provide, in the retail context, a mechanism for
electronic payment that does not rely upon accounts, and in the wholesale
context, a means of settlement that is more robust and less operationally
burdensome than present approaches.  It is not to create a substitute for bank
deposits, which would still be needed for economically important functions such
as fractional reserve banking, credit creation, and deposit insurance.  Neither
is it a replacement for cash, which offers a variety of benefits including
financial inclusion, operational robustness, and the assurance that a
transaction will complete without action on the part of third parties.  We
imagine that in practice, digital currency would be used primarily to
facilitate remittances that cannot be done using physical cash and that people
would not be more likely to be paid in digital currency in the future than they
would to be paid in cash today.

Nevertheless, we intend our proposed design to replicate some of the features
of cash.  Specifically, we seek to achieve the following properties:

\begin{enumerate}

\item\cz{Resistance to mass surveillance.}  Cash allows its bearers to transact
without fear that they will be profiled on the basis of their activities.  In
Section~\ref{ss:fraud}, we shall explicitly demonstrate that our design is
unlikely to increase the risk of fraud or AML/KYC violations relative to the
current system by comparing our proposed system to cash. In fact, we suspect
that it will lead to the opposite effect, given the possibility for the use of
digital analysis tools in the cases of regulated activities wherein adherence
to certain specific compliance rules is required and analysis over regulated
institutions activities is helpful.

\item\cz{Transaction assurance.}  Cash allows its bearers to know that a
potential transaction will succeed without depending upon a custodial or third-party
relationship that might block, delay, or require verification for a transaction to take place.

\item\cz{Non-discrimination.}  Cash allows is bearers to know that their money
is as good as everyone else's, and specifically that its value is not
determined by the characteristics of the bearer.

\end{enumerate}

We imagine that many, but not necessarily all, ordinary people and businesses
would have bank accounts into which they would receive payments.  These bank
accounts would sometimes earn interest made possible by the credit creation
activities of the bank.  Banks would be able to exchange digital currency at
par for cash or central bank reserves and would not generally hold wallets
containing an equal amount of digital currency to match the size of their
deposits.  In the case of CBDC, banks would also be able to directly exchange
the digital currency for central bank reserves.  When an individual (or
business) asks to withdraw digital currency, the bank would furnish it, just as
it would furnish cash today.  The bank might have a limited amount of digital
currency on hand just as it might have a limited amount of cash on hand to
satisfy such withdrawal requests, and there would be limits on the size and
rate of such withdrawals just as there would be limits on the size and rate of
withdrawals of cash.  Once they have digital currency, individuals and
businesses could use it to make purchases or other payments, as an alternative
to account-based payment networks or bank transfers, and digital currency would
generally be received into wallets held by regulated MSBs, just as cash would
be.

\subsection{The Role of Distributed Ledger Technology}
\label{ss:dlt}

\begin{figure}
\begin{center}
\scalebox{0.8}{\hspace{-12pt}\begin{tikzpicture}[
    >=latex,
    node distance=3cm,
    font={\sf},
    auto
]
\ts
\tikzset{>={Latex[width=4mm,length=4mm]}}
\tikzstyle{noshape} = [
    rectangle,
    text width=14em,
    text centered,
    align=center,
    minimum width=8cm,
    minimum height=1cm
]
\def\smbwd{3cm}

\node (block1) at (0,0) [block, fill=orange!50] {Digital Money Systems};
\node (block2) at (-5,-3) [block, fill=magenta!50] {Token-Based};
\node (block3) at (5,-3) [block, fill=brown!50] {Account-Based};
\node (block4) at (-8.33,-6) [block, fill=blue!30] {Distributed Ledgers};
\node (block5) at (-1.67,-6) [block, fill=red!30] {Centralised Ledgers};

\node (desc2) at (5, -4.5) [noshape] {pairwise reconciliation and reporting;
anonymous transactions are not possible.};

\node (desc4) at (-8.33, -7.75) [noshape] {ex ante distributed consensus process;
record of transactions is synchronised among participants.};

\node (desc5) at (-1.67, -7.5) [noshape] {validity of each transaction is
determined by a particular arbiter.};

\draw[->, ultra thick] (block1) -| (block2);
\draw[->, ultra thick] (block1) -| (block3);
\draw[->, ultra thick] (block2) -| (block4);
\draw[->, ultra thick] (block2) -| (block5);

\end{tikzpicture}}

\caption{\cz{Taxonomy of Digital Money Systems.}}

\label{f:taxonomy}
\end{center}
\end{figure}
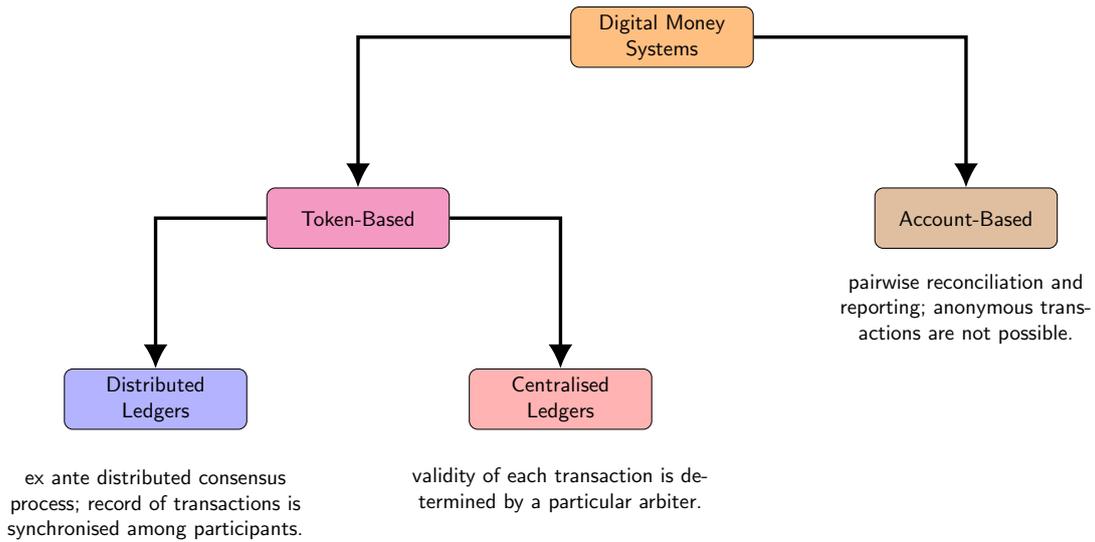

\noindent DLT offers a way to share responsibility for rulemaking among a set
of peers.  In the context of digital currency, DLT would provide transparency
to the operation and rules of the system by restricting (at a technical level)
what any single actor, including the central bank as well as government
regulators, can decide unilterally.  Such transparency complements and does not
substitute for regulatory oversight.

Although it is theoretically possible to build digital currency infrastructure
using centralised technology, we argue that for it to be successful, it will be
necessary for the infrastructure to use a distributed ledger.\footnote{This
should not be interpreted to mean that the infrastructure must provide for or
allow peer-to-peer transactions among users.}  First, consider the taxonomy of
digital money systems shown in Figure~\ref{f:taxonomy}.  Digital money systems
include CBDC and FDC.  The first question to ask is whether we need a system
based on tokens rather than a system based on accounts.  There are several
benefits to using a token-based system, including substantially reducing the
overhead associated with pairwise reconciliation and regulatory reporting.
Most importantly, however, any system based upon accounts cannot offer privacy,
since its design would necessarily require resolvable account identifiers that
can ultimately be used to determine both counterparties to any transaction.
Therefore, we must recognise that preservation of a token-based medium of
exchange is necessary to the public interest, increases welfare, and maintains
the critical nature of cash while providing to central banks and governments
the assurance and risk assessment tools that are afforded to digital payment
infrastructure platforms.

There are some important questions to ask about a token-based design, including
whether we need the tokens to be issued by the central bank directly, or by
other institutions (``stablecoins''), or whether the tokens can operate
entirely outside the institutional milieu (``cryptocurrency'').  However, let
us first understand why a distributed ledger is necessary.  Token-based systems
can be centralised, relying upon a specific arbiter to handle disputes about
the validity of each transaction (possibly with a different arbiter for
different transactions), or they can be decentralised, using a distributed
ledger to validate each transaction \textit{ex ante} via a consensus process.

Specifically, we consider the question of who the system operators would be.
In the case of CBDC, for example, although we assume that the central bank
would be responsible for the design and issuance of CBDC tokens, we do not make
the same assumption about the responsibility for the operation of a transaction
infrastructure or payment system, which historically has generally been
operated by private-sector organisations.  As mentioned earlier, systems for
payments, clearing, and settlement are often a collaborative
effort~\cite{bis2012,bis2012a}.  Indeed, modern digital payments infrastructure
based on bank deposits depends upon a variety of actors, and we imagine that
digital payments infrastructure based on CBDC would do so as well.  The
responsibility to manage and safeguard the value of currency is not the same as
the responsibility to manage and oversee transactions, and the responsibility
to supervise payment systems is not the same as the responsibility to operate
them.  A design that externalises responsibility for the operation of a
transaction infrastructure supporting CBDC is not incompatible with the
operational role of a central bank in using CBDC to create money and implement
monetary policy.

For reasons that we shall articulate in this section, we argue that a
token-based solution based on distributed ledger technology is required.  In
our view, the benefits of distributed ledger technology broadly fall into three
categories, all of which relate to the scope for errors, system compromise, and
potential liability arising from exogenous or endogenous risk scenarios.  We
believe that each of these benefits is indispensable and that all of them are
necessary for the system to succeed:

\begin{enumerate}

\item\cz{Eliminating the direct costs and risks associated with operating a
live system with a role as master or the capacity to arbitrate.}  Because its
database is centrally managed, a centralised ledger would necessarily rely upon
some central operator that would have an operational role in the transactions.
This operational role would have the following three implications.  First, the
central operator would carry administrative responsibility, including the
responsibility to guarantee system reliability on a technical level and handle
any exceptions and disputes on both a technical and human level.  Second,
because the central operator would be positioned to influence transactions, it
would incur the cost of ensuring that transactions are carried out as expected
as well as the risk of being accused of negligence or malice whether or not
they are carried out as expected.  Third, because the central operator
unilaterally determines what is allowed and what is not, it might be accused of
failing to follow the established rules.

\item\cz{Preventing unilateral action on the part of a single actor or group.}
Following the argument of Michael Siliski~\cite{siliski2018}, the administrator
of a centralised ledger could ban certain users or favour some users over
others; implicitly or explicitly charge a toll to those who use the system;
tamper with the official record of transactions; change the rules at any time;
or cause it to stop functioning without warning.

\item\cz{Creating process transparency and accountability for system
operators.}  Because the administrator of a centralised ledger can make
unilateral decisions, there is no way for outside observers to know whether it
has carried out its responsibilities directly.  In particular, its management
of the ledger and the means by which other parties access the ledger are under
its exclusive control, and the administrator has no need to publicise its
interest in changing the protocol or ask others to accept its proposed changes.
With DLT, it is possible to implement \textit{sousveillance} by ensuring that
any changes to the rules are explicitly shared with private-sector operators.

\item\cz{Improving efficiency and service delivery through competition and
scope for innovation.}  Vesting accountability for system operation in
operators who are incentivised to perform would make it possible to achieve
important service delivery objectives, ranging from adoption in the first
instance to financial inclusion and non-discrimination, through private-sector
incentives (e.g.  supporting local banks) rather than top-down political
directives.

\end{enumerate}

Each of these advantages of distributed ledger technology relates to the scope
for errors, system compromise, and potential liability arising from exogenous
or endogenous risk factors surrounding a central authority.  DLT makes it
possible to assign responsibility for transactions to the MSBs themselves.
Specifically, an MSB is responsible for each transaction that it writes to the
ledger, and the DLT can be used to create a (potentially) immutable record
binding each transaction to the corresponding MSB that submitted it, without
the need for a central actor would to be responsible for individual
transactions.

\subsection{System Design Overview}

Our design for DVC is based on the approach described as an
\textit{institutionally mediated private value exchange} by Goodell and
Aste~\cite{goodell2019}, which we elaborate here and further build upon.  This
proposal uses DLT for payments, as motivated by reasons articulated in
Section~\ref{ss:dlt}.

We envision a \textit{permissioned} distributed ledger architecture wherein the
participants would be regulated MSBs.  MSBs would include banks, other
financial businesses such as foreign exchange services and wire transfer
services, as well as certain non-financial businesses such as post
offices~\cite{bis2012} as well.  The \textit{permissioned} DLT design would
support efficient consensus mechanisms such as Practical Byzantine Fault
Tolerance~\cite{castro1999}, with performance that can be compared to popular
payment networks.  In particular, Ripple has demonstrated that its network can
reliably process 1,500 transactions per second~\cite{ripple2019}.  Although the
popular payment network operator Visa asserts that its system can handle over
65,000 transactions per second~\cite{visa2018}, its actual throughput is not
more than 1,700 transactions per second~\cite{visa2020}.  For this reason, we
anticipate that it will be possible for a digital currency solution to achieve
the necessary throughput requirement without additional innovation.

We assume that the only parties that could commit transactions to the ledger
and participate in consensus would be MSBs, which would be regulated entities.
The ledger entries would be available for all participants to see, and we
imagine that certain non-participants such as regulators and law enforcement
would receive updates from the MSBs that would allow them to maintain copies of
the ledger directly, such that they would not need to query any particular MSB
with specific requests for information.  Although the ledger entries themselves
would generally not contain metadata concerning the counterparties, the MSB
that submitted each transaction would be known to authorities, and it is
assumed that MSBs would maintain records of the transactions, including
transaction size and whatever information they have about the counterparties
even if it is limited, and that authorities would have access to such records.

Another important feature of our proposed architecture is \textit{privacy by
design}.  Although we argue that data protection is no substitute for privacy
(see Section~\ref{ss:landscape}), Ulrich Bindseil notes that ``others will
argue that a more proportionate solution would consist in a sufficient
protection of electronic payments data''~\cite{bindseil2020}.  In the case of
our proposed design, we might imagine that because the entire network is
operated by regulated MSBs, some people might recommend creating a ``master
key'' or other exceptional access mechanisms to allow an authority to break the
anonymity of retail DVC users.  The temptation to build exceptional access
mechanisms should be resisted, with appreciation for the history of such
arguments~\cite{abelson1997,abelson2015,benaloh2018} and subsequent
acknowledgement by policymakers in Europe and
America~\cite{hr5823,thomson2016}, who have repeatedly cited their potential
for abuse as well as their intrinsic security vulnerabilities.  Ultimately,
substituting data protection for privacy would create a dragnet for law-abiding
retail DVC users conducting legitimate activities, and it will never be
possible for a data collector to prove that data have not been subject to
analysis.  To force people to use a system that relies on data protection is to
attempt to manufacture trust, which is impossible; trust must be earned.
Furthermore, criminals and those with privilege will have a variety of options,
including but not limited to proxies, cryptocurrencies, and identity theft,
available to them as ``outside solutions'' in the event that lawmakers attempt
to force them into transparency.

Unlike designs that contain exceptional access mechanisms that allow
authorities to trace the counterparties to every transaction and therefore do
not achieve anonymity at all, our approach actually seeks to deliver true but
``partial'' anonymity, wherein the counterparties to a transaction can be
anonymous but all transactions are subject to control at the interface with the
MSB.  We believe that our design is unique in that it achieves both anonymity
and control by ensuring that all transactions involve a regulated actor but
without giving authorities (or insiders, attackers, and so on) the ability to
unmask the counterparties to transactions, either directly or via correlation
attacks.

To satisfy the requirement for privacy by design, we introduce the concept of a
\textit{private wallet}, which is software that interacts with the ledger via
an MSB that allows a retail DVC user to unlink her DVC tokens from any
meaningful information about her identity or the identity of any previous
owners of the tokens.  Specifically, a transaction in which a fungible token
flows from a private wallet to an MSB reveals no meaningful information about
the history of the token or its owner.  To support private wallets, a DVC
system must incorporate certain privacy-enhancing technology of the sort used
by privacy-enabling cryptocurrencies such as Zcash and Monero.  There are at
least two possible approaches~\cite{iso-tr-23244}:

\begin{enumerate}

\item \cz{Stealth addresses, Pedersen commitments, and ring signatures.}
Stealth addresses, which obscure public keys by deriving them separately from
private keys~\cite{courtois2017}, deliver privacy protection to the receiver of
value~\cite{iso-tr-23244}.  Pedersen commitments, which obscure the amounts
transacted to anyone other than the transacting
parties~\cite{pedersen1991,vanwirdum2016}, remove transaction metadata from the
ledger records~\cite{iso-tr-23244}.  Ring signatures, which allow signed
messages to be attributable to ``a set of possible signers without revealing
which member actually produced the signature''~\cite{rivest2001}, deliver
privacy protection to the sender of value~\cite{iso-tr-23244}.

\item \cz{Zero-knowledge proofs.} Zero-knowledge proofs ``allow one party to
prove to another party that a statement is true without revealing any
information apart from the fact that the statement is
true''~\cite{iso-tr-23244} and can potentially be used to protect all of the
transaction metadata~\cite{iso-tr-23244}.  Non-interactive approaches to
zero-knowledge proofs such as ZK-STARKs deliver significant performance
advantages over their interactive alternatives~\cite{ben-sasson2018}, and based
upon their measured performance~\cite{ben-sasson2018,guan2019,zcash-sapling},
we anticipate that such operations will be fast enough to suffice for
point-of-sale or e-commerce transactions.

\end{enumerate}

\begin{figure}[t]
\begin{center}
\begin{tikzpicture}[>=latex, node distance=3cm, font={\sf \small}, auto]\ts
\node (box1) at (8,0.2) [box, minimum width=2.8cm, minimum height=3.4cm] {};
\node (r1) at (0,0) [noshape, text width=4em] {
    \scalebox{0.8}{\includegraphics{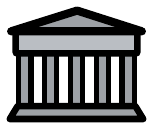}}
};
\node (s1) at (0.3,-0.2) [box, color=white, fill=magenta] {
    MSB
};
\node (c1) at (0,-4) [noshape, text width=4em] {
    \scalebox{0.8}{\includegraphics{images/primary_bank.pdf}}
};
\node (o1) at (0,4) [noshape, text width=4em] {
    \scalebox{1.0}{\includegraphics{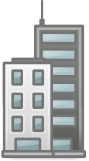}}
};
\node (r2) at (4,0) [noshape, text width=4em] {
    \scalebox{0.8}{\includegraphics{images/primary_bank.pdf}}
};
\node (s2) at (4.3,-0.2) [box, color=white, fill=magenta] {
    MSB
};
\node (c2) at (4,-4) [noshape, text width=4em] {
    \scalebox{0.8}{\includegraphics{images/primary_bank.pdf}}
};
\node (o2) at (4,4) [noshape, text width=4em] {
    \scalebox{0.5}{\includegraphics{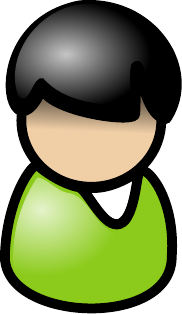}}
};
\node (r3) at (8,0) [noshape, text width=4em] {
    \scalebox{0.06}{\includegraphics{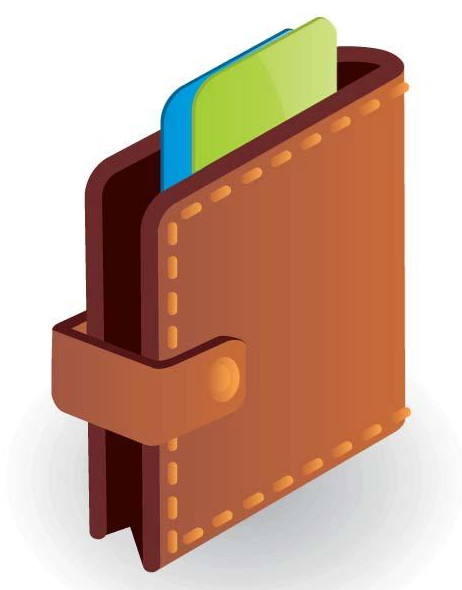}}
};
\node (o3) at (8,4) [noshape, text width=4em] {
    \scalebox{0.5}{\includegraphics{images/people-juliane-krug-08a.pdf}}
};
\node (r4) at (12,0) [noshape, text width=4em] {
    \scalebox{0.8}{\includegraphics{images/primary_bank.pdf}}
};
\node (s4) at (12.3,-0.2) [box, color=white, fill=magenta] {
    MSB
};
\node (c4) at (12,-4) [noshape, text width=4em] {
    \scalebox{0.8}{\includegraphics{images/primary_bank.pdf}}
};
\node (o4) at (12,4) [noshape, text width=4em] {
    \scalebox{1.0}{\includegraphics{images/office-towers.pdf}}
};
\node (m1) at (0.8,0.8) [noshape] {
    \scalebox{0.06}{\includegraphics{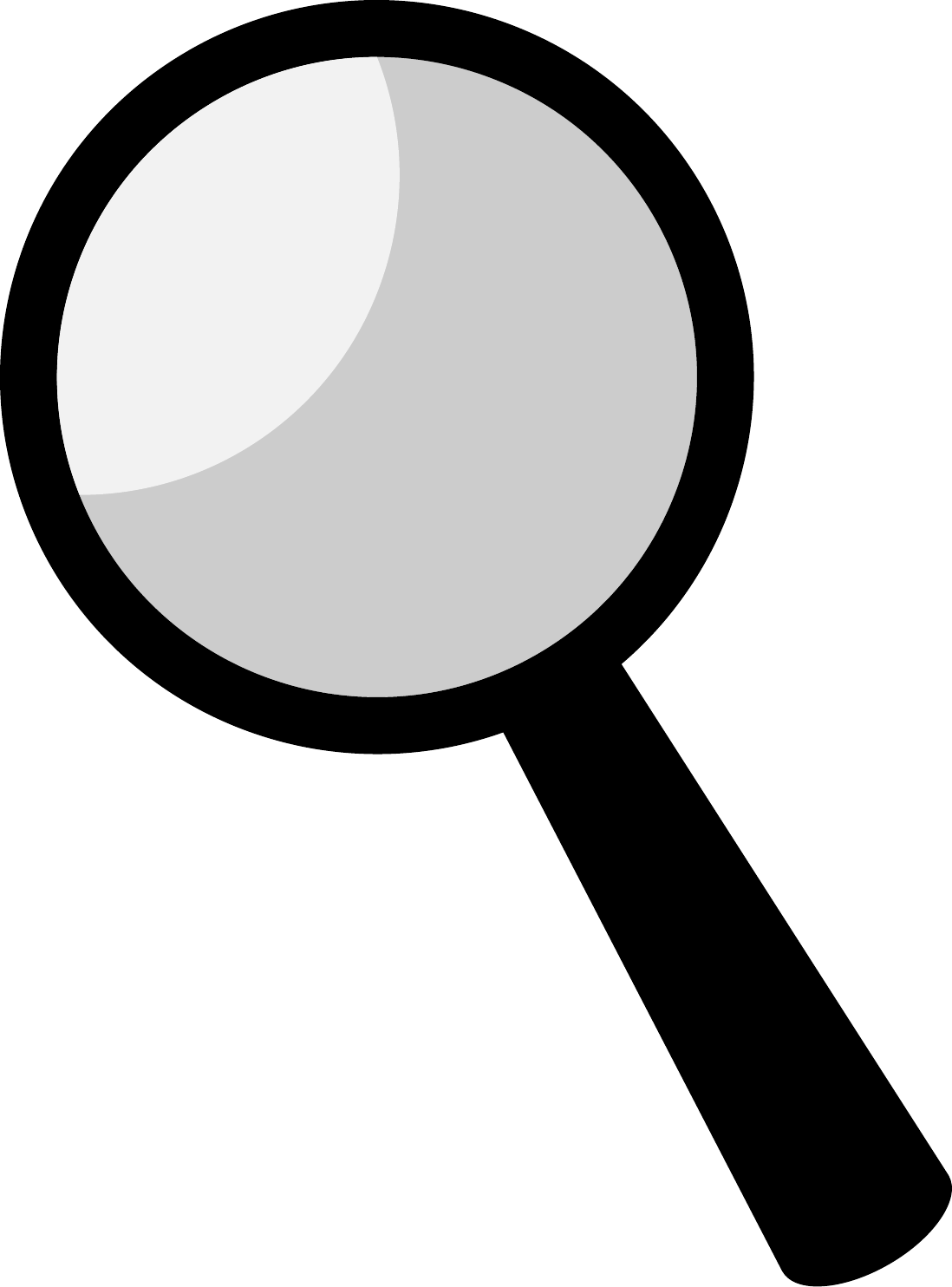}}
};
\node (m2) at (4.8,0.8) [noshape] {
    \scalebox{0.06}{\includegraphics{images/office-glass-magnify.pdf}}
};
\node (m2) at (12.8,0.8) [noshape] {
    \scalebox{0.06}{\includegraphics{images/office-glass-magnify.pdf}}
};
\node (m1) at (8.8,0.8) [noshape] {
    \scalebox{0.06}{\includegraphics{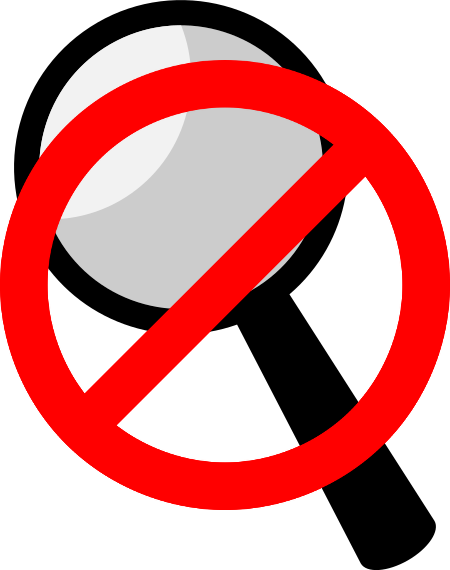}}
};

\draw[->, line width=0.5mm, align=center] (o1) -- node[below] {
    Payment\\(e.g. payroll)
} (o2);
\draw[->, line width=0.5mm, align=center] (o3) -- node[below] {
    Payment\\(e.g. purchase)
} (o4);
\draw[->, dashed, line width=0.5mm] (r1) -- node[below] {
    \textbf{\textit{\Huge \pounds}}
} (r2);
\draw[->, dashed, line width=0.5mm] (0, -3.4) -- node[right] {
    \textbf{\textit{\Huge \pounds}}
} (0, -1.2);
\draw[->, line width=0.5mm] (r2) -- node[below] {
    \textbf{\textit{\Huge \pounds}}
} (r3);
\draw[->, dashed, line width=0.5mm] (4, -3.4) -- node[right] {
    \textbf{\textit{\Huge \pounds}}
} (4, -1.2);
\draw[->, line width=0.5mm] (r3) -- node[below] {
    \textbf{\textit{\Huge \pounds}}
}(r4);
\draw[->, dashed, line width=0.5mm] (12, -1.2) -- node[right] {
    \textbf{\textit{\Huge \pounds}}
} (12, -3.4);
\node (d1) at (0,3) [nos] {Business A};
\node (d2) at (4,3) [nos] {Individual B};
\node (d3) at (8,3) [nos] {Individual B};
\node (d4) at (12,3) [nos] {Business C};
\draw[<->, line width=0.5mm, align=center] (0, 2.8) -- node[sloped, above] {
    account
} (0, 0.6);
\draw[<->, line width=0.5mm, align=center] (4, 2.8) -- node[sloped, above] {
    account
} (4, 0.6);
\draw[<->, line width=0.5mm, align=center] (8, 2.8) -- (8, 0.6);
\draw[<->, line width=0.5mm, align=center] (12, 2.8) -- node[sloped, above] {
    account
} (12, 0.6);
\node (e1) at (0,-0.8) [nos] {Bank A};
\node (e2) at (4,-0.8) [nos] {Bank B};
\node (e3) at (8,-0.8) [nos] {private wallet};
\node (e4) at (12,-0.8) [nos] {Bank C};
\node (f1) at (0,-4.8) [nos] {central bank};
\node (f2) at (4,-4.8) [nos] {central bank};
\node (f4) at (12,-4.8) [nos] {central bank};

\end{tikzpicture}

\caption{\cz{Schematic representation of a typical user engagement
lifecycle.}  Individual B first receives an ordinary payment from Business A,
which holds an account with Bank A, into her account with Bank B.
Next, the individual asks Bank B to withdraw CBDC from Bank B
into her private wallet.  On-ledger transactions of CBDC are represented by the
Pound Sterling symbol (\pounds).  (If Bank B had not received the CBDC
directly from Bank A along with the payment, then it might source the
CBDC from its own holdings, or it might receive the CBDC from the central bank
in exchange for cash or reserves.)  Finally, the individual makes a payment to
Business C, which Business C receives into its account with Bank C,
which then has the option to return the CBDC to the central bank in exchange
for cash or reserves.}

\label{f:pevept}
\end{center}
\end{figure}

\subsection{User Engagement Lifecycle}

Figure~\ref{f:pevept} depicts a typical user engagement lifecycle with CBDC,
which we anticipate would be a typical use case for DVC.  This user has a bank
account and receives an ordinary payment via bank transfer into her account.
Then, the user asks her bank to withdraw CBDC, which takes the form of a set of
tokens that are effectively transferred to her private wallet via a set of
transactions to different, unlinkable addresses that her bank publishes to the
ledger.  Later, the user approaches a merchant (or other service provider,
either in-person or online, with a bank account that is configured to receive
CBDC.  Using her private wallet, the user interacts with point-of-sale software
operated by the business, which brokers an interaction between her private
wallet and the merchant's bank wherein the bank publishes a set of transactions
to the ledger indicating a transfer of CBDC from the user's private wallet to
the bank, credits the merchant's account, and informs the merchant that the
transaction was processed successfully.  The privacy features of the ledger
design and the private wallet software ensure that the user does not reveal
anything about her identity or the history of her tokens in the course of the
transaction that can be used to identify her or profile her behaviour.  More
generally, we envision that a retail user of digital currency would receive it
via one of four mechanisms:

\begin{figure}[t]
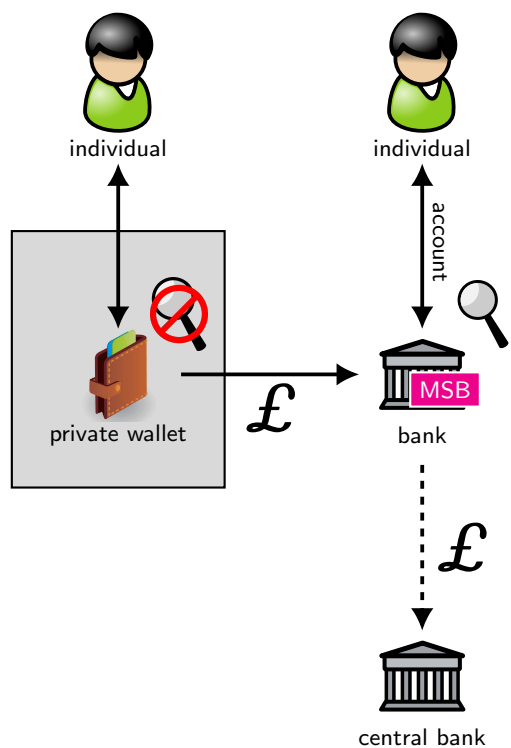

\begin{center}
\begin{tikzpicture}[>=latex, node distance=3cm, font={\sf \small}, auto]\ts
\node (box1) at (8,0.2) [box, minimum width=2.8cm, minimum height=3.4cm] {};
\node (r3) at (8,0) [noshape, text width=4em] {
    \scalebox{0.06}{\includegraphics{images/wallet-vector-xp.png}}
};
\node (o3) at (8,4) [noshape, text width=4em] {
    \scalebox{0.5}{\includegraphics{images/people-juliane-krug-08a.pdf}}
};
\node (r4) at (12,0) [noshape, text width=4em] {
    \scalebox{0.8}{\includegraphics{images/primary_bank.pdf}}
};
\node (s4) at (12.3,-0.2) [box, color=white, fill=magenta] {
    MSB
};
\node (c4) at (12,-4) [noshape, text width=4em] {
    \scalebox{0.8}{\includegraphics{images/primary_bank.pdf}}
};
\node (o4) at (12,4) [noshape, text width=4em] {
    \scalebox{0.5}{\includegraphics{images/people-juliane-krug-08a.pdf}}
};
\node (m2) at (12.8,0.8) [noshape] {
    \scalebox{0.06}{\includegraphics{images/office-glass-magnify.pdf}}
};
\node (m1) at (8.8,0.8) [noshape] {
    \scalebox{0.06}{\includegraphics{images/office-glass-magnify-no.png}}
};
\draw[->, line width=0.5mm] (r3) -- node[below] {
    \textbf{\textit{\Huge \pounds}}
}(r4);
\draw[->, dashed, line width=0.5mm] (12, -1.2) -- node[right] {
    \textbf{\textit{\Huge \pounds}}
} (12, -3.4);
\node (d3) at (8,3) [nos] {individual};
\node (d4) at (12,3) [nos] {individual};
\draw[<->, line width=0.5mm, align=center] (8, 2.8) -- (8, 0.6);
\draw[<->, line width=0.5mm, align=center] (12, 2.8) -- node[sloped, above] {
    account
} (12, 0.6);
\node (e3) at (8,-0.8) [nos] {private wallet};
\node (e4) at (12,-0.8) [nos] {bank};
\node (f4) at (12,-4.8) [nos] {central bank};

\end{tikzpicture}

\caption{\cz{Schematic representation of a user depositing CBDC into a bank
account.} Retail users would be permitted to deposit funds into their own
accounts, possibly subject to certain limits or additional checks in the event
that such deposits are frequent or large.}

\label{f:pevdeposit}
\end{center}
\end{figure}

\begin{figure}[t]
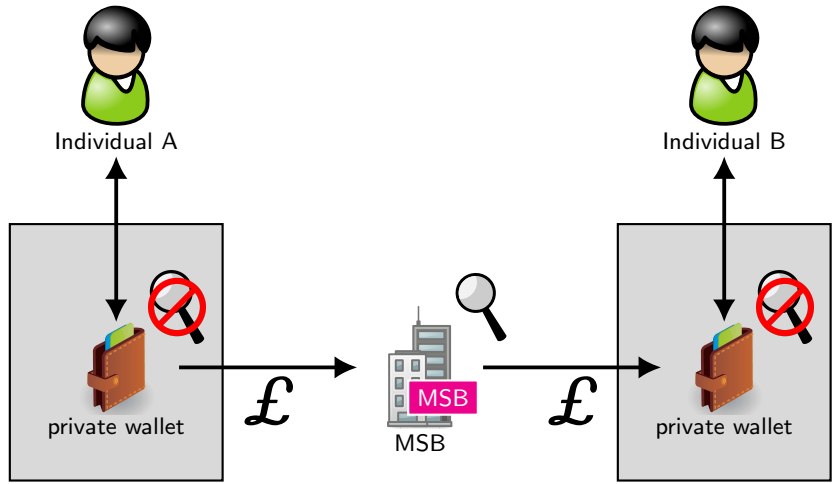

\begin{center}
\begin{tikzpicture}[>=latex, node distance=3cm, font={\sf \small}, auto]\ts
\node (box1) at (0,0.2) [box, minimum width=2.8cm, minimum height=3.4cm] {};
\node (box1) at (8,0.2) [box, minimum width=2.8cm, minimum height=3.4cm] {};
\node (r1) at (0,0) [noshape, text width=4em] {
    \scalebox{0.06}{\includegraphics{images/wallet-vector-xp.png}}
};
\node (o1) at (0,4) [noshape, text width=4em] {
    \scalebox{0.5}{\includegraphics{images/people-juliane-krug-08a.pdf}}
};
\node (r2) at (4,0) [noshape, text width=4em] {
    \scalebox{1.0}{\includegraphics{images/office-towers.pdf}}
};
\node (s2) at (4.3,-0.4) [box, color=white, fill=magenta] {
    MSB
};
\node (r3) at (8,0) [noshape, text width=4em] {
    \scalebox{0.06}{\includegraphics{images/wallet-vector-xp.png}}
};
\node (o3) at (8,4) [noshape, text width=4em] {
    \scalebox{0.5}{\includegraphics{images/people-juliane-krug-08a.pdf}}
};
\node (m1) at (0.8,0.8) [noshape] {
    \scalebox{0.06}{\includegraphics{images/office-glass-magnify-no.png}}
};
\node (m2) at (4.8,0.8) [noshape] {
    \scalebox{0.06}{\includegraphics{images/office-glass-magnify.pdf}}
};
\node (m3) at (8.8,0.8) [noshape] {
    \scalebox{0.06}{\includegraphics{images/office-glass-magnify-no.png}}
};

\draw[->, line width=0.5mm] (r1) -- node[below] {
    \textbf{\textit{\Huge \pounds}}
}(r2);
\draw[->, line width=0.5mm] (r2) -- node[below] {
    \textbf{\textit{\Huge \pounds}}
}(r3);
\node (d1) at (0,3) [nos] {Individual A};
\node (d3) at (8,3) [nos] {Individual B};
\draw[<->, line width=0.5mm, align=center] (0, 2.8) -- (0, 0.6);
\draw[<->, line width=0.5mm, align=center] (8, 2.8) -- (8, 0.6);
\node (e1) at (0,-0.8) [nos] {private wallet};
\node (e2) at (4,-1) [nos] {MSB};
\node (e3) at (8,-0.8) [nos] {private wallet};

\end{tikzpicture}

\caption{\cz{Schematic representation of a mediated transaction between
consumers.}  Retail CBDC users wishing to transact with each other via their
private wallets must transact via a regulated institution or a regulated
business with an account with a regulated institution.  The institution creates
on-ledger transactions from the private wallet of one retail CBDC user and to
the private wallet of another retail CBDC user without creating accounts for
the retail CBDC users.}

\label{f:pevmed}
\end{center}
\end{figure}

\begin{figure}[t]
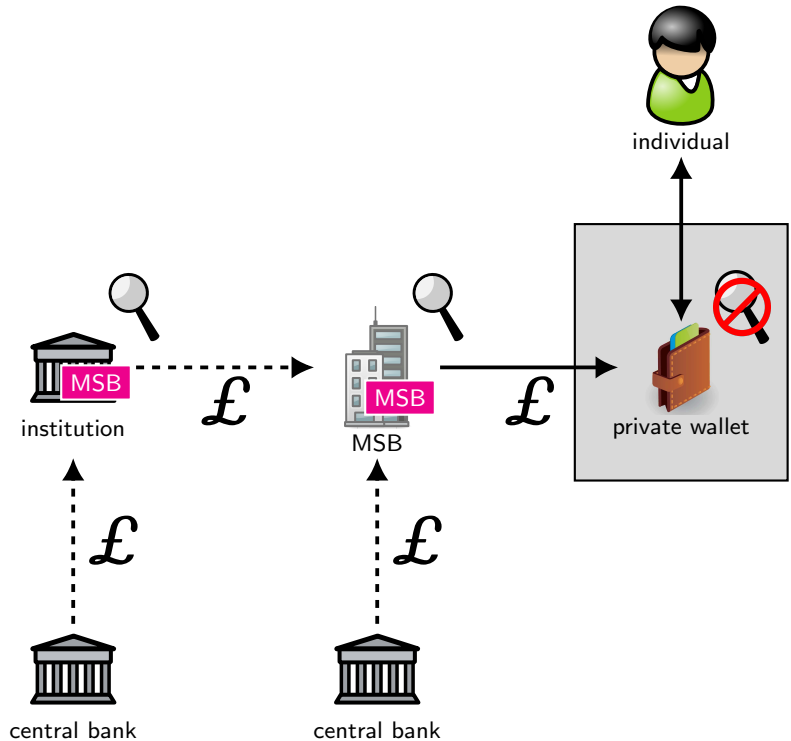

\begin{center}
\begin{tikzpicture}[>=latex, node distance=3cm, font={\sf \small}, auto]\ts
\node (box1) at (8,0.2) [box, minimum width=2.8cm, minimum height=3.4cm] {};
\node (r2) at (4,0) [noshape, text width=4em] {
    \scalebox{1.0}{\includegraphics{images/office-towers.pdf}}
};
\node (s2) at (4.3,-0.4) [box, color=white, fill=magenta] {
    MSB
};
\node (c0) at (0,-4) [noshape, text width=4em] {
    \scalebox{0.8}{\includegraphics{images/primary_bank.pdf}}
};
\node (c1) at (0,0) [noshape, text width=4em] {
    \scalebox{0.8}{\includegraphics{images/primary_bank.pdf}}
};
\node (r3) at (8,0) [noshape, text width=4em] {
    \scalebox{0.06}{\includegraphics{images/wallet-vector-xp.png}}
};
\node (c2) at (4,-4) [noshape, text width=4em] {
    \scalebox{0.8}{\includegraphics{images/primary_bank.pdf}}
};
\node (o3) at (8,4) [noshape, text width=4em] {
    \scalebox{0.5}{\includegraphics{images/people-juliane-krug-08a.pdf}}
};
\node (m1) at (0.8,0.8) [noshape] {
    \scalebox{0.06}{\includegraphics{images/office-glass-magnify.pdf}}
};
\node (m2) at (4.8,0.8) [noshape] {
    \scalebox{0.06}{\includegraphics{images/office-glass-magnify.pdf}}
};
\node (m3) at (8.8,0.8) [noshape] {
    \scalebox{0.06}{\includegraphics{images/office-glass-magnify-no.png}}
};

\draw[->, dashed, line width=0.5mm] (c1) -- node[below] {
    \textbf{\textit{\Huge \pounds}}
}(r2);
\draw[->, line width=0.5mm] (r2) -- node[below] {
    \textbf{\textit{\Huge \pounds}}
}(r3);
\node (d3) at (8,3) [nos] {individual};
\draw[<->, line width=0.5mm, align=center] (8, 2.8) -- (8, 0.6);
\node (e1) at (0,-0.8) [nos] {institution};
\node (s1) at (0.3,-0.2) [box, color=white, fill=magenta] {
    MSB
};
\node (e2) at (4,-1) [nos] {MSB};
\node (e3) at (8,-0.8) [nos] {private wallet};
\draw[->, dashed, line width=0.5mm] (0, -3.4) -- node[right] {
    \textbf{\textit{\Huge \pounds}}
} (0, -1.2);
\draw[->, dashed, line width=0.5mm] (4, -3.4) -- node[right] {
    \textbf{\textit{\Huge \pounds}}
} (4, -1.2);
\node (f1) at (0,-4.8) [nos] {central bank};
\node (f2) at (4,-4.8) [nos] {central bank};

\end{tikzpicture}

\caption{\cz{Schematic representation of a disbursement to a retail user with a
private wallet.}  This example shows how a retail user might claim CBDC that
she is eligible to receive, either directly from the central bank or from an
institution such as the State treasury or a private-sector bank.  The user
would identify herself to a regulated MSB, which would carry out the requisite
compliance checks.  (FDC could be received similarly, except without the
involvement of the central bank.)}

\label{f:pevstimulus}
\end{center}
\end{figure}

\begin{enumerate}

\item \cz{Via an exchange of money from an account with an MSB into digital
currency.}  We stipulate that an individual or business with an account with an
MSB could opt to \textit{withdraw} digital currency from the account into a
private wallet.  Digital currency held by a retail user in the user's private
wallet would be like cash.  Because it is not held by an MSB, it would not be
invested and it would not earn true interest.  (In Section~\ref{s:analysis}, we
suggest a mechanism by which governments can incentivise or penalise the asset
itself, but this would not be ``true'' interest and would not serve the same
purpose.)  Similarly, an individual or business with an account with an MSB
could opt to \textit{deposit} digital currency from a private wallet into an
account, reversing the process, as shown in Figure~\ref{f:pevdeposit}.

\item \cz{As a recipient of digital currency from an external source,
received into an account with an MSB.}  In this case, the user would be the
recipient of a digital currency payment.  The sender of the payment might be
known, for example if it is an account with an MSB, or it might be unknown,
specifically if it is a private wallet.

\item \cz{As a recipient of digital currency from an external source, received
into a private wallet.}  Any transaction in which a private wallet receives
digital currency from an external source must be mediated by an MSB, so the key
difference between this mode of receiving digital currency and a withdrawal
from the user's own account is that in this case the recipient does not have
(or is not using) an account with the MSB.  This form of transaction is
illustrated in Figure~\ref{f:pevmed}.  We imagine that there would be certain
legal requirements, such as transaction limits or a requirement for the
recipient to provide positive identification documents to a human clerk, that
would govern the role of the MSB in such transactions.  We also imagine that
this process could be particularly useful as a means to deliver government
payments (for economic stimulus or for other reasons) to retail users without
bank accounts, as illustrated in Figure~\ref{f:pevstimulus}.

\item \cz{Via an exchange of physical cash into digital currency.}  The
transaction in which physical cash is converted to digital currency would be
facilitated by an MSB, subject to appropriate rules, just as in the case that
digital currency is received directly from an external source.  For example,
the MSB might be required to ask for information concerning the origin of the
cash if the amount exceeds a certain threshold.

\end{enumerate}

Note that retail bank accounts are not generally expected to hold CBDC on
behalf of a particular user, any more than retail bank accounts would hold cash
on behalf of a particular user.  A bank would swap CBDC for central bank
reserves from time to time, and vice-versa, with the expectation that the bank
would furnish CBDC to its retail customers, subject to limits on the size and
rate of withdrawals.

Note also that the messages on the ledger are published by regulated financial
institutions.  This is an important feature of the system design: all
transactions on the ledger must be published by a regulated MSB, and because
the ledger is operated entirely by regulated MSBs, private actors cannot
exchange value directly between their private wallets.  At the same time, the
private wallets offer a layer of indirection wherein MSBs would not be able to
identify the \textit{counterparties} to the transactions involving private
wallets.  Banks might need to know their customers, but merchants generally do
not.  Furthermore, a merchant's bank does not need to know the merchant's
customers, and a merchant's customer's bank does not need to know about the
merchant or its bank at all.  For instances wherein merchants really do need to
know their customers, the reason is generally about the substance of the
relationship rather than the mechanism of the payment, and identification of
this sort should be handled outside the payment system.

By providing a mechanism by which no single organisation or group would be able
to build a profile of any individual's transactions in the system, the use of a
distributed ledger achieves an essential requirement of the design.  In
addition to our previously stated requirement that transactions into and out of
the private wallets would be protected by mechanisms such as stealth addresses
or zero-knowledge proofs to disentangle the outflows from the inflows,
individuals would be expected to use their private wallets to transact with
many different counterparties, interacting with the MSBs chosen by their
counterparties and not with the MSBs from which their private wallets were
initially funded.

Figure~\ref{f:pevmed} depicts the mechanism by which individuals would transact
from one private wallet to another.  They must first identify a regulated MSB
to process the transaction onto the ledger, perhaps in exchange for a small
fee.  The MSB would process a set of transactions from the first private wallet
to the MSB and from the MSB to the second private wallet.  An MSB could provide
a similar service for an individual exchanging CBDC for cash or vice-versa.
Presumably, the MSB would gather whatever information is needed from its
customers to satisfy compliance requirements, although we imagine that strong
client identification, such as what might conform to the FATF
recommendations~\cite{fatf-recommendations}, could be waived for transactions
that take place in-person and are sufficiently small.  In the case of small
online transactions between two persons, we imagine that an attribute-backed
credential indicating that either the sender or the receiver is eligible to
transact might be sufficient~\cite{goodell2019a}.  Finally, some MSBs could
provide token-mixing services for retail DVC users who had accidentally exposed
metadata about the tokens in their private wallets.

Concerning the hypothetical stimulus described in Figure~\ref{f:pevstimulus},
we note that if a government intends to make stimulus payments to a specific
set of eligible individuals,\footnote{The government could do the same for
businesses, if desired.} notwithstanding the possibility that this set might
include all citizens or residents, then it could refer to each such individual
using a unique taxpayer identification number.  Then, the government could ask
each eligible party to specify a bank account, current account, or wallet into
which to deposit the funds.  This approach might work in many cases, although
it might not work for eligible individuals or busineses without bank accounts.
To address the gap, the government could ask eligible parties to identify
themselves to a qualified MSB for verification, for example a post office, that
would be able to carry out the required identification procedures to determine
whether the prospective recipient has the right to make a claim associated with
a particular taxpayer identification number.  Once this is done, the MSB could
enter a transaction that delivers the digital currency to the individual's
private wallet directly, avoiding the need for a bank account. We propose that
each of these options could be provided to both individuals and businesses.

\subsection{Security Considerations}

Since digital currencies generally rely upon the use and management of
sensitive cryptographic information such as keys, we recognise that a digital
currency that allows users to hold tokens outside of the protection of an
account with a financial institution would also introduce responsibility on the
part of users to manage the security of those tokens.  Users have a range of
possible options at their disposal, including encrypted devices with one-factor
or two-factor authentication, third-party custodial services, single-use
physical tokens as an alternative to wallet software for their general-purpose
devices, and simply choosing to limit the amount of digital currency that they
hold at any moment.  We suggest that all of these approaches could be useful,
and as with many financial decisions, the best choice would be a function of
the preferences and risk profile of each individual user.

We imagine that an individual might share the private cryptographic information
(e.g. a private key that can be used to initiate a transaction) associated with
digital currency with another individual, thereby allowing the other individual
to transact it on her behalf.  We do not consider that such an exchange of
information would constitute a payment, since there is nothing intrinsic to the
system that would stop the first party from spending the digital currency
before the second party has a chance to do so.  It would be appropriate to
characterise such an exchange as a ``promise of payment'' rather than a payment
itself, similar to providing a post-dated cheque, and there is no mechanism to
prevent people from making promises to each other.  Once an individual or
business is in possession of digital currency, the ways to dispose of the
digital currency are the inverses of the methods to acquire it.

\subsection{System Governance}

Because privacy-enhancing technologies require vigilance~\cite{zimmermann1991},
MSBs and the broader community must commit to maintain, audit, challenge, and
improve the technology underpinning the privacy features of this design as part
of an ongoing effort~\cite{goodell2019}.  Such maintenance implies establishing
a process for security updates as well as updates to accommodate new technology
and features as needed.  The transparency afforded by the use of DLT can
provide the basis by which the broader community can observe and analyse the
operation of the system, including any changes to its regular functioning, to
ensure that transacting parties remain protected against technologically
sophisticated adversaries with an interest in de-anonymising the DVC users for
the purpose of profiling them.

Ultimately, whoever controls the code that the system relies upon to operate,
controls the operation of the system.  By analogy, consider the role of
developer communities in handling ledger-related disputes in cryptocurrency
communities~\cite{wong2016}.  For this reason, a centralised developer
community could certainly negate the benefit of a decentralised ledger.  This
implies that each independent participant in the system should establish its
own rigorous procedure for accepting changes to the code, most likely including
internal code review and security analysis, whether or not participants share
the same code base, and it might be necessary for this process to be subject to
public oversight as well.  Such procedures for internal and external oversight
should involve a broad security community with diverse allegiances, and in
particular, care must be taken to ensure that it will be possible to make
timely changes to address emerging problems\footnote{Including but not limited
to shutdowns and partial shutdowns.} while protecting both users and system
operators from the possibility that backdoors or other vulnerabilities might be
introduced in haste.  This is no simple task, although the work of the security
community in free software projects such as Debian~\cite{debian-security}
demonstrate that the combination of deep oversight and timely changes is
possible, and established procedures for the operation of trading networks such
as the National Market System in the United States~\cite{nms-changes},
demonstrate that such changes can be undertaken in a co-regulatory context,
with formal proposals by regulators, as well.

From the standpoint of DVC, platform governance and decision-making
predominantly relates to authenticating and thereby allowing transactions. Our
proposal, as summarised in Table~\ref{t:table} contends that the infrastructure
would be operated by the private sector and may be exclusively operated by the
private sector. We envisage that there should be no fewer than five MSBs for a
pilot, and no fewer than about twenty MSBs for robust operation. The approval
of transactions takes place through consensus across the infrastructure
operators of the platform. However, the ability to formally become an
infrastructure operator and MSB \textit{pro tanto} requires the approval of the
local regulator, however it is regulated.  We assume in this context the
central bank is responsible for overseeing clearing and settlement
activities.\footnote{For example, in the case of the United Kingdom it may be
through joint oversight between the Prudentical Regulatory Authority (PRA) and
the Financial Conduct Authority (FCA) for matters related to conduct.}

\section{Analysis}
\label{s:analysis}

\begin{figure}
\begin{center}
\scalebox{0.6}{\hspace{-12pt}\begin{tikzpicture}[
    >=latex,
    node distance=3cm,
    font={\sf},
    auto
]
\ts
\tikzset{>={Latex[width=4mm,length=4mm]}}
\tikzstyle{noshape} = [
    rectangle,
    text width=14em,
    text centered,
    align=center,
    minimum width=8cm,
    minimum height=1cm
]
\def\smbwd{3cm}

\node (box2) at (3.9,6) [
    boxy,
    minimum height=12cm,
    minimum width=9.75cm,
    fill=red!15
] {};
\node (block4) at (6.5,5) [block, fill=orange!50] {Treasury Securities};
\node (block5) at (6.5,1) [block, fill=orange!50] {Currency};
\node (box1) at (11,3) [boxy, minimum height=6cm, minimum width=4cm, fill=blue!20] {};
\node (box3) at (24.4,3) [boxy, minimum height=6cm, minimum width=4cm, fill=blue!20] {
    {\large \,}\\
    {\large \,}\\
    {\large \,}\\
    {\large Liquidity}\\
    {\large \,}\\
    {\large Demand}\\
    {\large \,}\\
    {\large Spending}\\
    {\large \,}\\
    {\large Investment}\\
    {\large \,}\\
};
\node (block6) at (11,5) [block, fill=orange!50] {
    Fiscal Policy
};
\node (block7) at (11,1) [block, fill=orange!50] {
    Monetary Policy
};

\coordinate (cn) at (3.9, 8);
\draw[->, ultra thick] (cn) |- (block4);
\draw[->, ultra thick] (cn) |- (block5);
\draw (3.9,11) node[anchor=south]{\large Scalability}
  -- (1.5,7) node[anchor=north]{\large Privacy}
  -- (6.3,7) node[anchor=north]{\large Control}
  -- cycle [thick, fill=white];
\node (n) at (3.9,8) [noshape, minimum width=0pt] {\large Trilemma};
\coordinate (x) at (13.2,3);

\node (p1) at (11,6.3) [noshape] {\large Policy Tools};
\node (p2) at (24.4,6.3) [noshape] {\large Real Economy};

\draw[->, ultra thick] (block4) -- (block6);
\draw[->, ultra thick] (block5) -- (block7);
\draw[->, ultra thick] (box1) -- (box3);

\end{tikzpicture}}\\
\vspace{0.5em}
\cz{(a)\vspace{1.5em}} Government-Issued Assets and Policy.\\
\scalebox{0.6}{\hspace{-12pt}\begin{tikzpicture}[
    >=latex,
    node distance=3cm,
    font={\sf},
    auto
]
\ts
\tikzset{>={Latex[width=4mm,length=4mm]}}
\tikzstyle{noshape} = [
    rectangle,
    text width=14em,
    text centered,
    align=center,
    minimum width=8cm,
    minimum height=1cm
]
\def\smbwd{3cm}

\node (box2) at (3.9,6) [
    boxy,
    minimum height=12cm,
    minimum width=9.75cm,
    fill=red!15
] {};
\node (block1) at (0.5,-1) [block, fill=orange!50] {
    Digital Value Container (\textbf{DVC})
};
\node (block2) at (3,3) [block, fill=orange!50] {
    National DVC (\textbf{NDVC})\\
    \vspace{5pt} Transnational DVC (\textbf{TDVC})
};
\node (block3) at (3,-3) [block, fill=orange!50] {
    \textbf{Private DVC}
};
\node (block4) at (6.5,5) [block, fill=orange!50] {
    Fiscal Digital Currency (\textbf{FDC})
};
\node (block5) at (6.5,1) [block, fill=orange!50] {
    Central Bank Digital Currency (\textbf{CBDC})
};
\node (box1) at (11,3) [boxy, minimum height=6cm, minimum width=4cm, fill=blue!20] {};
\node (box3) at (24.4,3) [boxy, minimum height=6cm, minimum width=4cm, fill=blue!20] {
    {\large \,}\\
    {\large \,}\\
    {\large \,}\\
    {\large Liquidity}\\
    {\large \,}\\
    {\large Demand}\\
    {\large \,}\\
    {\large Spending}\\
    {\large \,}\\
    {\large Investment}\\
    {\large \,}\\
};
\node (block6) at (11,5) [block, fill=orange!50] {
    Fiscal Policy
};
\node (block7) at (11,1) [block, fill=orange!50] {
    Monetary Policy
};

\coordinate (cn) at (3.9, 8);
\node (block8) at (15.5,3) [block, fill=orange!50, text width=10em] {
    \textbf{Asset Features}\\
    (e.g. vintage, quality, stability in use)
};
\node (block9) at (19.9,3) [block, fill=orange!50, text width=10em] {
    \textbf{System Features}\\
    (e.g. programmability, privacy)
};
\draw[->, ultra thick] (cn) -| (block8);
\draw[->, ultra thick] (cn) -| (block9);
\draw (3.9,11) node[anchor=south]{\large Scalability}
  -- (1.5,7) node[anchor=north]{\large Privacy}
  -- (6.3,7) node[anchor=north]{\large Control}
  -- cycle [thick, fill=white];
\node (n) at (3.9,8) [noshape, minimum width=0pt] {\large Trilemma};
\node (n0) at (18,1.5) [noshape, minimum width=0pt] {\large Public-Private Coordination};
\coordinate (x) at (13.2,3);

\node (p1) at (11,6.3) [noshape] {\large Policy Tools};
\node (p2) at (24.4,6.3) [noshape] {\large Real Economy};

\draw[->, ultra thick] (block1) |- (block2);
\draw[->, ultra thick] (block1) |- (block3);
\draw[->, ultra thick] (block2) |- (block4);
\draw[->, ultra thick] (block2) |- (block5);
\draw[->, ultra thick] (block4) -- (block6);
\draw[->, ultra thick] (block5) -- (block7);
\draw[->, ultra thick] (box1) -- (block8);
\draw[->, ultra thick] (block8) -- (block9);
\draw[->, ultra thick] (block9) -- (box3);

\end{tikzpicture}}\\

\vspace{0.5em}
\cz{(b)} Government-Issued Assets and Policy with Digital Value Containers.

\caption{\cz{Facilitating economic activity and exchange with Digital Value
Containers.}}

\label{f:economic}
\end{center}
\end{figure}

Although it can accommodate CBDC, the digital currency system we propose is
actually a ``value container''~\cite{pilkington2016}, which we extend to
potentially represent a plethora of different assets and their underlying
infrastructure, including but not limited to central bank or government assets.
The system can be adapted to work with both fiat-based CBDC issued by central
banks as well as with asset-based FDC issued by governments, and for this
reason it can facilitate more effective monetary and fiscal policy.  Our
proposal responds directly to concerns that have been introduced or exacerbated
by the SARS-CoV-2 crisis as a signal of the innovation that should have been
taking place.  In particular, our proposed system can be used as a flexible
platform for implementing a variety of economic stimulus policies including
streamlined issuance of currency by the central bank, QE through asset
purchases, and direct payment or allocation of credits to individuals and
non-financial businesses.  Figure~\ref{f:economic} offers an illustration of
how digital value containers can enhance the policy toolkit available to State
actors.

\subsection{Remuneration and Incentives}

Our CBDC design model allows for all CBDCs  (or some part of them) to bear
interest-like features and (dis)incentives to encourage either spending or
saving.  These features can be applied more in general also to FDCs and could
be done through ``vintages'' where tokens are marked according to their time of
issue allowing the CBDC or FDC to mimic the effects of remuneration.\footnote{In
the case of vintages, holders are able to verify the vintage of each token they
hold.}  It is critical to note here that we are not referring to
(dis)incentives on accounts directly, but rather associated instead to the
tokens themselves and which are set by the issuer, subject to legal terms that
would define the number of tokens in the current vintage that the central bank
or other issuing authority would provide in exchange for one token of an
earlier vintage.  We anticipate that these terms would be communicated at the
time of issuance.

We anticipate that an MSB would be legally required to send only tokens of the
latest vintage, which is to say that tokens received by an MSB are effectively
non-transferable, except to the issuer.  We further anticipate that an MSB
would be legally required to accept tokens of any vintage on behalf of a retail
user, although if the token is not of the most recent vintage, then the value
received by the recipient would be determined by the legal terms for the
treatment of tokens of the particular vintage.  The MSB would not necessarily
be required to immediately exchange with the issuer any old-vintage tokens that
it receives.  However, if it were to do so, then the number of new-vintage
tokens that it receives from the issuer from executing the exchange would be
equal to the value that it would record as having been received.  If the issuer
is the central bank, as in the case of CBDC, then an arrangement would be
required whereby the net cost incurred or surplus received by the central bank
would flow from the national treasury as a subsidy or would flow to the
national treasury as a tax.

Once the decision to implement vintages is made, the designer can experiment
with different design options. For example, different classes of vintages can
be introduced and either subsidies or taxes can be set or levied on the
exchange across one variant to another, or through penalties and rewards.
Vintages will not generally be usable to distinguish one token of CBDC or FDC
from another, as a timestamp would, instead, those of a particular vintage
would be fungible with one another. Each vintage would encompass all CBDC or
FDC issued within a particular time period that is, preferably and plausibly,
set to be long enough that it is realistic for most retail users to generally
be expected to have spent all of the tokens they receive at the start of the
vintage period before the end of the vintage; for example, it would be possible
to have a few vintage periods per year or one per year.

Consider the case in which there are several live vintages during a time period
$T$.  Although we have stated that tokens from different vintages will be
fungible for one another, they are not perfect substitutes for one another.
Suppose that a government implements a policy to discourage hoarding of tokens
by establishing legal terms implying that the value of a token issued during
period $T$ will have a value less than a token issued in period $T+1$.  The
tokens from the two vintages remain nonetheless fungible in being denominated
in the local tender, although for retail users to verify the value of their
tokens in terms of the most recent vintage will require clear communication
during issuance.  Similarly, wallets would generally be capable of calculating
the value of a token at a particular point in time as it decays or appreciates
in value according to the legal terms of its vintage.

We suggest that it would generally be prudent to require the schedule, type,
and format of rewards and/or penalties over the lifetime of a CBDC or FDC
within a particular vintage to be generally fixed at the time of issuance,
although this is not strictly necessary and would be subject to the policy
objectives at hand.  We do not envisage that rewards or penalties would apply
to a particular vintage until after one full vintage period following the
issuance of that vintage but it would be possible. Under this assumption, an
issued DVC in vintage period $T$ would be exchangeable at an MSB for bank
deposits at par with cash or bank deposits, for the entirety of vintage periods
$T$ and $T+1$.

To penalise long-term retail users and encourage spending, a government might
impose a penalty at a certain rate $r$ that would apply to all tokens of that
vintage at the start of vintage period $T+2$ and each vintage period
thereafter, meaning that its value following the penalty would be multiplied by
$1-r$, compounding every year thereafter until period $T+n$, where $n$ is
chosen by the issuer, after which time the token is declared worthless.  On the
other hand, to reward long-term retail users and encourage saving, a government
could set as a feature a reward at a certain rate $r$ that would apply to all
tokens of that vintage at the start of vintage period $T+2$ and each vintage
period thereafter, meaning that its value following the reward would be
multiplied by $1+r$, compounding every year thereafter until period $T+n$, and
expiring sometime thereafter. However, it is important to note that if the
reward were fixed to a specific annual rate at issue, it would appear to
operate like a Guaranteed Investment Certificate (GIC) or Certificate of
Deposit (CD).  If no expiration date were specified, it might be likened to a
consolidated annuity.  We have advocated transparent communication of
dis(incentives) to provide clarity to users and holders but it could be
conceivable that rates of decay or reward could be accelerated, delayed, or
extended in certain circumstances. Moreover, were the policy rates and market
rates to decrease, there would be an incentive to hold the CBDC or FDC, driving
holders to view and treat the tokens as an investment.  As we previously
described, we would envisage that issuers would develop appropriate modelling
such that were there to be a reward for holding tokens of some vintage $T$,
there would also be tokens of other vintages that could, if necessary, be
delayed or extended to ensure appropriate policy alignment across policy rates
while ensuring that the overall policy serves the public interest.

In principle, the DVC issuer could also issue a ``special vintage'' of DVC
during some period $T$ that can only be transacted by retail users before some
time $T+n$, with the purpose of stimulating spending, with rules preventing
MSBs from allowing retail users to exchange this special vintage for bank
deposits or cash.  After time $T+n$, retail users might have a short grace
period in which to exchange their DVC tokens for cash or bank deposits with a
penalty, after which time the tokens would expire worthless.

The ability to impose positive incentives such as through subsidies on CBDC
would increase demand for it and lead to an outflow from bank notes and
possibly bank deposits into CBDC, whereas were it to be negative or through
disincentives such as taxes, as a measure taken to escape a liquidity trap, it
would decrease the demand for CBDC and lead to an outflow from cash as a safer
price stable medium of exchange.  To address this major issue, together with
the possibility of arbitrage across instrument types, would require policy
makers to make the right DVC design choices. For example, if the overnight rate
were less than the remuneration $r$ of CBDCs, MSBs would hold CBDC, which would
ultimately decrease the yield on CBDC and increase the overnight rate. If the
overnight rate were greater than that on CBDC, then MSBs would invest into
reserves which would increase the yield on CBDC. Therefore, it would be
pragmatic for the interest rate on CBDC and overnight right on reserves to be
equivalent, and which would ultimately help set other market rates within the
economy through the appropriate (dis)incentives such as taxes and subsidies in
a particular vintage~\cite{27}.

Were the CBDC or FDC to carry with it (dis)incentives in some way, it would
become a priority monetary and fiscal policy instrument through its
programmability in the conventional monetary and fiscal policy toolbox, because
it would affect household spending and firm investment saving decisions either
(i) directly by means of funds in their respective wallets through
taxes/penalties or rewards/subsidies to the DVC asset itself, whereby real-time
policy could become a reality; (ii) indirectly due the remuneration of the CBDC
leading to the setting of the lower limit for the rate on bank deposits; and
(iii) modelled to achieve directly, the same intended effects of setting of the
policy rate. Our CBDC platform can be used as a tool to achieve these goals
seamlessly with the flexibility to alter features directly, in an automated way
and immediately.

\subsection{Impact on Liquidity}

The issuance and use of a CBDC could also become a useful tool for central
banks in managing aggregate liquidity. For example, were CBDC to be widely held
and adopted for use, it could lead to a shift in aggregate liquidity, which
refers to the assets being used and exchanged and which carry a liquidity
premium~\cite{39}. Under certain models, a CBDC would lead to efficient
exchange, particularly given that it is a low cost medium of exchange and has a
stable unit of account, and particularly in the case wherein the digital
currency (as we propose it) is being used in a broad range of decentralised
transactions, and allows for monetary policy transmission channels on trading
activity to be strengthened. The central bank would have at its disposal
certain capabilities in controlling the supply and price of CBDC, including
through the use of (dis)incentives to generate a higher liquidity or lower
premium in CBDC and in bank deposits, subject to where investment frictions
exist in a much more targeted way~\cite{39}.  Moreover, CBDC can be used as
intraday liquidity by its holders, whereas liquidity-absorbing instruments
cannot achieve the same effect. At present, there are few short-term money
market instruments that inherently combine the creditworthiness and the
liquidity that a CBDC (or DVC, more generally) could potentially provide. CBDC,
therefore, could play an important deterrent role against liquidity shocks.

One possible concern about CBDC is that individuals might run from bank
deposits to CBDC during a financial crisis.  Although such a run is
conceivable, we argue that it is no more likely with our proposed system for
CBDC than it is with cash.  Specifically, we imagine that individuals would be
subject to limits on their withdrawals of CBDC from their bank accounts, just
as they are subject to limits on their withdrawals of cash.  If a run were
underway, its pace would be limited by such limits, and in principle, the
government could even ask banks to impose tighter limits or to disallow
withdrawals from banks entirely in the event of an emergency. Moreover, if the
government chooses to guarantee bank deposits up to an amount, then the other
benefits afforded by such deposits coupled with that guarantee would
disincentivise such a run.  In other instances the cost-benefit and risk-reward
profile would require more specific analysis on a jurisdiction by jurisdiction
basis.  Because we recognise significant utility for bank deposits even in the
presence of DVC, we suggest that DVC would be be complementary to deposits and
that banks would play a fundamental role in the issuance and storage of DVC.

\subsection{Impact on the Financial Industry}

The most direct impact of our approach to digital currency on the financial
industry involves risk management, on several levels.  By improving the speed
of settlement, digital currency can be used to facilitate liquidity risk
management among financial institutions. Digital currency can also be used to
address systemic risk, both explicitly, by offering regulators a view into
substantially every transaction, as well as implicitly, by offering governments
a tool to implement stimulus while controlling the aggregate leverage in the
system.

Considering that, in general, DLT offers a promising risk-mitigation tool
\cite{TascaMorini}, our design relies on a DLT network operated by MSBs and
other private-sector institutions rather than a centralised ledger run by a
single public (or private\footnote{Like in all the stablecoin solutions.})
organisation. As such, our approach addresses a variety of risks associated
with relying upon a central arbiter: (1) technical risks associated with
availability, reliability, and maintenance; (2) risks associated with trust and
operational transparency; and (3) financial and legal risks.  Our approach also
allows the private sector to operate the infrastructure for retail payments,
clearing, and settlement, while allowing government regulators to oversee the
system at an organisational level.  Because we imagine that digital currency
will complement rather than substitute for bank deposits, our approach
leverages the role of commercial banks without forcibly decreasing their
balance sheets.  In particular, because we believe that the main purpose of DVC
tokens will be to facilitate electronic payments rather than to serve as a
long-term store of value, we do not anticipate that the balance sheets of
central banks will increase significantly as a result of its introduction.

\subsection{Impact on Fraud and Tax Evasion}
\label{ss:fraud}

We imagine that a rigorous compliance regime will govern the behaviour of MSBs
and the relationships they have with their customers.  We assume that banks in
particular will have requirements for strong customer identification, and other
MSBs such as wire transfer firms, currency exchanges, and post offices will
face a combination of transaction limitations and procedures for identification
and authorisation.  We assume that authorities will be able to see every
transaction that takes place as well as the specific MSB that creates that
transaction, and we also assume that authorities will have access to the
records that the MSBs are required to maintain concerning the transactions they
facilitate.

Nevertheless, because our system allows a measure of true anonymity, it does
not provide a way to reveal the identities of both counterparties to
authorities.  In particular, even if authorities have all of the records, some
transactions will have private wallets as a counterparty, just as some cash
transactions have anonymous counterparties.  Although authorities might know
all of the retail users and their history of digital currency withdrawals, they
will not be able to link a private wallet to a specific retail user.  Recall
that retail users will be able to withdraw digital currency from an MSB in the
same manner that they would withdraw cash from a bank or ATM, with similar
limits and restrictions.  Retail users would be able to spend digital currency
the same way that they would be able to spend cash, making purchases with
vendors who are also subject to limits and restrictions as well as profiling by
their financial institutions, and who know that their receipt of tokens will be
monitored by authorities.  Authorities would know who had recently withdrawn
digital currency into a private wallet just as they would know who had recently
withdrawn cash, and they would also know who had recently received digital
currency from a private wallet.  However, it would not be possible to use the
digital currency to link a specific recipient of cash to a specific
counterparty that had made a withdrawal.  We argue that this property of cash
is necessary and fundamental to protect retail users from profiling and
manipulation by adversaries and other powerful interests including private
sector participants.  Furthermore, revealing mutual counterparty information
for every transaction would divert the onus of fraud detection to law
enforcement agencies, effectively increasing their burden, while well-motivated
criminals would still be able to use proxies or compromised accounts to achieve
their objectives, even if every transaction were fully transparent.

To manage fraud, our system design takes a different approach that is oriented
toward control mechanisms and transaction analytics rather than counterparty
profiling.  Because every transaction involves a regulated financial
intermediary that would presumably be bound by AML/KYC regulations, there is a
clear path to investigating every transaction effectively.  Authorities would
be positioned to ensure that holders of accounts that take payments from
private wallets adhere to certain rules and restrictions, including but not
limited to tax monitoring.  The records from such accounts, combined with the
auditable ledger entries generated by the DLT system, could enable real-time
collection of data concerning taxable income that could support reconciliation
and compliance efforts.  Because all of the retail payments involving digital
currency would ultimately use the same ledger, identification of anomalous
behaviour, such as a merchant supplying an invalid destination account for
remittances from private wallets, would be more straightforward than in the
current system, and real-time automated compliance would be more readily
achievable.  Such detection could even be done in real-time not only by
authorities but also by customers, thus reducing the likelihood that it would
occur in the first instance.

It is worth considering whether safely storing large amounts of physical cash
would be more or less costly than storing large amounts of digital currency.
In principle, digital currency can be stored cheaply online, although the
attack surface of online systems might have important weaknesses, and the
longevity of offline digital media has limits.  Note that security safes are
generally priced as a function of the value, not the storage cost, of what is
inside.  In addition, the use of vintages can explicitly penalise the
accumulation of large stashes of digital currency in a manner that is hard to
replicate with physical cash.

It is also worth considering whether criminal organisations might exchange
private keys rather than entering transactions on the ledger as a way to avoid
interacting with MSBs.  Our view is that sharing a private key is equivalent to
sharing the ability to spend money that can only be spent once, effectively
constituting a promise, otherwise as transferring posession in the case of a
private wallet.  Criminals can exchange promises by a variety of private or
offline methods even in the absence of a privacy-respecting payment system.  At
one level, it is impossible to monitor or restrict such exchanges of promises,
but at another level, exchanges of this sort would require a high degree of
\textit{a priori} trust to succeed, and we submit that transitive trust
relationships would generally degrade rapidly across successive transactions.
Meanwhile, attempts to spend the same token twice can be easily detected, and
potentially investigated, by authorities at the time of the transaction. In our
view, the utility derived from the privacy preserving nature of a payment
infrastructure warrants a trade-off, however, the trade-off is substantially
limited given the added capability available to law enforcement and the
mechanisms that may be instituted, coupled with the fact that would there to be
nefarious actors and activities, those activities could take place in a variety
of ways and media, and they are not more effectively enabled by our system.

\subsection{Comparison to Alternative Approaches}

\begin{table}[t]
\begin{center}

\sf
\begin{tabular}{|L{9.3cm}|p{\ccol}p{\ccol}p{\ccol}p{\ccol}p{\ccol}p{\ccol}p{\ccol}p{\ccol}p{\ccol}|}\hline
& \rotatebox{90}{Goodell, Al-Nakib, Tasca}
& \rotatebox{90}{R3~\cite{r3-cbdc}}
& \rotatebox{90}{Bank of England~\cite{boe2020}}
& \rotatebox{90}{Sveriges Riksbank~\cite{riksbank}}
& \rotatebox{90}{Adrian and Mancini-Griffoli (IMF)~\cite{adrian2019}\,\,\,}
& \rotatebox{90}{Bordo and Levin~\cite{16}}
& \rotatebox{90}{ConsenSys~\cite{consensys}}
& \rotatebox{90}{Zhang ``Synthetic CBDC'' (IMF)~\cite{zhang2020}}
& \rotatebox{90}{Auer and B\"ohme (BIS)~\cite{auer2020a}}\\\hline
Can hold value outside an account                                   & \CIRCLE & \Circle & \Circle & \Circle & \Circle & \Circle & \Circle & \Circle & \Circle \\
DLT system                                                          & \CIRCLE & \CIRCLE & \Circle & \CIRCLE & \Circle & \Circle & \CIRCLE & \Circle & \Circle \\
No central gatekeeper for transactions                              & \CIRCLE & \CIRCLE & \Circle & \CIRCLE & \CIRCLE & \Circle & \CIRCLE & \CIRCLE & \Circle \\
Can be operated exclusively by private, independent actors          & \CIRCLE & \CIRCLE & \Circle & \CIRCLE & \CIRCLE & \Circle & \CIRCLE & \CIRCLE & \Circle \\
State manages issuance and destruction                              & \CIRCLE & \CIRCLE & \CIRCLE & \CIRCLE & \CIRCLE & \CIRCLE & \CIRCLE & \Circle & \CIRCLE \\
Retail users do not hold accounts with the central bank             & \CIRCLE & \CIRCLE & \CIRCLE & \Circle & \CIRCLE & \Circle & \Circle & \CIRCLE & \CIRCLE \\
True privacy (in contrast to data protection)                       & \CIRCLE & \Circle & \Circle & \Circle & \Circle & \Circle & \Circle & \Circle & \Circle \\
All transactions are on-ledger                                      & \CIRCLE & \CIRCLE & \CIRCLE & \CIRCLE & \Circle & \CIRCLE & \CIRCLE & \Circle & \CIRCLE \\
All transactions require a regulated intermediary                   & \CIRCLE & \CIRCLE & \CIRCLE & \Circle & \CIRCLE & \CIRCLE & \Circle & \CIRCLE & \CIRCLE \\
Supports tax or subsidy directly to the asset                       & \CIRCLE & \CIRCLE & \Circle & \CIRCLE & \CIRCLE & \CIRCLE & \CIRCLE & \Circle & \Circle \\
Supports arbitrary fungible assets and (dis)incentives to bearers   & \CIRCLE & \Circle & \Circle & \Circle & \Circle & \Circle & \Circle & \Circle & \Circle \\
Intermediaries can include non-financial institutions               & \CIRCLE & \Circle & \Circle & \Circle & \Circle & \Circle & \CIRCLE & \Circle & \Circle \\
\hline\end{tabular}
\rm

\caption{\cz{Comparison of features among proposed retail digital currency architectures.}}

\label{t:table}
\vspace{-1em}
\end{center}
\end{table}

Table~\ref{t:table} offers a
comparison of the main design features.  The features of our design that
contrast with many of the prevailing CBDC design proposals include, but are not
limited to, the following:

\begin{enumerate}

\item\cz{Retail users can hold digital assets outside accounts.}  Most of the
existing proposals assume that digital assets would be always held by
intermediaries.  In contrast, our proposal empowers retail users with the
ability to truly control the assets they hold and choose custodians, when
applicable, on their own terms.

\item\cz{No central bank accounts for individuals and non-financial
businesses.} In our view, requiring central bank accounts would introduce new
costs, weaknesses, and security vulnerabilities.  It would result in the
central bank taking responsibility for actions commonly performed by the
private sector in many countries, and it would negate the benefits of using
tokens rather than accounts.  A team led by Jes\'us Fern\'andez-Villaverde
observed that many proponents of CBDC such as Bordo and Levin~\cite{16} assume
that central banks would disintermediate commercial intermediaries and that in
many cases this possibility is touted as a benefit of
CBDC~\cite{fernandez2020}.  However, their analysis formalises a trade-off
between avoiding bank runs and delivering optimal allocation of
capital~\cite{fernandez2020}, underscoring a key role of commercial banks in
bearing risk that, in our view, should not be undermined.

\item\cz{A purpose-built domestic, retail payment system.}  The requirement to
support cross-border or wholesale payments is intentionally not included in our
design.  Our proposal is designed specifically to meet the requirements for a
domestic, retail payment system, which we believe differ significantly from the
requirements for a cross-border or wholesale payment system.

\item\cz{True, verifiable privacy for retail users.}  Data protection is not
the same as privacy, and our proposal does not rely upon third-party trust or
data protection for their transaction metadata.  Some proposals include
``anonymity vouchers'' that would be usable for a limited time in
accounts-based digital currency systems~\cite{r3-cbdc,dgen}.  We do not believe
that such approaches would be effective, not only because of the dangers
associated with reducing the anonymity set to specific intervals but also
because of the attacks on anonymity that will always be possible if value is to
be transferred from one regulated account directly to another.

\item\cz{No new digital identity systems.}  Our system does not require any
special identity systems beyond those that are already used by MSBs and
private-sector banks.  In particular, it does not require a system-wide
identity infrastructure of any kind, and it also explicitly allows individuals
to make payments from their private wallets without revealing their identities.

\item\cz{No new real-time operational infrastructure managed by central
authorities.}  Our proposed system can be operated exclusively by private,
independent actors without relying upon a central actor to operate any specific
part of the infrastructure.  The distributed ledger makes it possible to assign
responsibility for most transactions to the MSBs, not the central bank.  An MSB
is responsible for each transaction that it writes to the ledger, and the DLT
can be used to create a (potentially) immutable record binding every
transaction to the corresponding MSB that submitted it.  We understand that the
central bank is not responsible for individual transactions.

\end{enumerate}

\section{Recommendations}

In an effort to address the current crisis, central banks and governments have
had to deploy a variety of conventional and unconventional policies and tools.
The community is aware that innovative solutions are needed both to address the
current crisis, which is markedly different than the last one in 2008-2011,
and, most importantly, to prepare for future crises.  Along this thread, a
growing community of academics and policymakers is actively discussing the
opportunity to empower central banks with a CBDC.  As a result, different
design models have been proposed so far in the literature.

We believe that all the models proposed so far fail to meet important design
criteria that have been summarised in Table 1. In particular, we show that
other concurrent CBDC design proposals omit certain design features that have
an impact on critical areas of welfare-generating characteristics, as well as
governance and financial implications.  The proposal that we have articulated
addresses these essential requirements directly and does not compromise.

At the same time, our proposal is also much broader than a version of digital
cash issued by a central bank. Instead, it encapsulates a generalised
underlying infrastructure that enables a greater and more direct variety of
tools to effect monetary and fiscal policy.  Much like a connected railway
system, our DVC proposal provides the underlying basis for programming a large
variety of assets in new ways, supporting their deployment and distribution,
and securing the privacy of users without disrupting existing market
structures. Whether it is used for CBDC or FDC, sovereign debt or collateral,
or other fungible and non-fungible assets, the underlying infrastructure is a
necessary next step to provide the appropriate framework to better equip
government policymakers to address systemic issues in the real economy. In
other words, our DVC proposal is complementary both to the current two-tiered
banking structure and to the array of existing monetary and fiscal policy tools
and their mechanisms.

We conclude by highlighting two final important design features that make our
model unique.  First, our proposal uses a DLT-based settlement system that is
overseen by State actors but operated entirely by private, independent actors.
Second, it aims to enhance the welfare and safety of users by employing
\textit{privacy by design} without compromising the core risk analysis capacity
in which policymakers would be interested if DVC were used to implement a CBDC
system.

\section*{Acknowledgements}

We thank Professor Tomaso Aste for his continued support for our project, we
thank Larry Wall of the Federal Reserve Bank of Atlanta for his valuable
feedback, and we acknowledge the support of the Centre for Blockchain
Technologies at University College London.  Geoff Goodell also acknowledges the
Centre for Technology and Global Affairs at the University of Oxford as well as
the European Commission for the FinTech project (H2020-ICT-2018-2 825215).

\sf

\noindent All icons and clipart images are available at
publicdomainvectors.org, with the exception of the wallet icon, which is
available at vectorportal.com.

\end{document}